%% file: deep_materials_cag_arxiv.tex
\documentclass[times,twocolumn,final]{elsarticle}
\usepackage{cag}
\usepackage{framed,multirow}

\usepackage{amssymb}
\usepackage{latexsym}

\usepackage{url}
\usepackage{xcolor}
\definecolor{newcolor}{rgb}{.8,.349,.1}

\usepackage{hyperref}

\usepackage[switch,pagewise]{lineno} %Required by command \linenumbers below

\journal{Computers \& Graphics}
\usepackage[utf8]{inputenc}
\usepackage[english]{babel}
\usepackage{amsmath}
\usepackage{amsfonts}
\usepackage{wrapfig}
\usepackage{lipsum}
\usepackage{placeins}
\usepackage{multirow}
\usepackage{hyperref}
\usepackage{graphbox}
\usepackage[normalem]{ulem}
\usepackage{rotating}
\usepackage{graphicx, subcaption}
\usepackage[linesnumbered,ruled, vlined,commentsnumbered]{algorithm2e}
\usepackage{algorithmic}
\usepackage{enumitem}
\usepackage{tikz}
\usetikzlibrary{calc,spy}
\tikzstyle{closeup} = [
opacity=1.0,          % opacity of both windows
height=2cm,         %  height of large window
width=2cm,          %   width of large window
]
\definecolor{myBorder0}{rgb}{0.332,0.268,1.0}
\definecolor{myBorder1}{rgb}{0.274,0.8498,0.3224}
\definecolor{myBorder2}{rgb}{1,0.758,0.274}
\definecolor{myBorder3}{rgb}{0.932,0.268,0.3574}
\tikzstyle{largewindow} = [myBorder0, line width=0.40mm]
\tikzstyle{smallwindow} = [myBorder0,line width=0.80mm]
\tikzstyle{largewindow2} = [myBorder1, line width=0.40mm]
\tikzstyle{smallwindow2} = [myBorder1,line width=0.80mm]
\tikzstyle{largewindow3} = [myBorder2, line width=0.40mm]
\tikzstyle{smallwindow3} = [myBorder2,line width=0.80mm]
\tikzstyle{largewindow4} = [myBorder3, line width=0.40mm]
\tikzstyle{smallwindow4} = [myBorder3,line width=0.80mm]

\author[1]{Siddhant \snm{Prakash}\corref{cor1}}
\emailauthor{siddhant.prakash@inria.fr}{Siddhant Prakash}
    
\author[1]{Gilles \snm{Rainer}}
\emailauthor{gilles.rainer@inria.fr}{Gilles Rainer}

\author[1]{Adrien \snm{Bousseau}}
\emailauthor{adrien.bousseau@inria.fr}{Adrien Bousseau}

\author[1]{George \snm{Drettakis}}
\emailauthor{George.Drettakis@inria.fr}{George Drettakis}

\address[1]{Universit\'{e} C\^{o}te d'Azur and Inria, Sophia-Antipolis, 06902, France}

\received{18 July 2022}
\accepted{28 September 2022}
\availableonline{10 October 2022}
\definecolor{todoCol}{rgb}{1,0.5,0.5}
\newcommand{\todo}  [1] { \textcolor{todoCol}{\textbf{TODO: }{#1}}}
\newcommand{\TODO}  [1] { \textcolor{todoCol}{\textbf{TODO: }{#1}}}
\newcommand{\SP}[1]{\textcolor{blue}{\textbf{SP: #1}}}
\newcommand{\GR}[1]{\textcolor{green}{\textbf{GR: #1}}}
\newcommand{\GD}[1]{\textcolor{cyan}{\textbf{GD: #1}}}
\newcommand{\AB}[1]{\textcolor{green}{\textbf{AB: #1}}}
\newcommand{\NEW}[1]{\textcolor{red}{#1}}
\newcommand{\CK}[1]{#1}

\renewcommand{\todo}  [1] { }
\renewcommand{\TODO}  [1] { }
\renewcommand{\SP}[1]{ }
\renewcommand{\GR}[1]{ }
\renewcommand{\AB}[1]{ }
\renewcommand{\GD}[1]{ }
\renewcommand{\NEW}[1]{#1}

\makeatother
\AtBeginDocument{%
  \providecommand\BibTeX{{%
    \normalfont B\kern-0.5em{\scshape i\kern-0.25em b}\kern-0.8em\TeX}}}

\verso{Author's Version}

\begin{document}

\begin{frontmatter}
\title{Deep scene-scale material estimation from multi-view indoor captures}


%\keywords{datasets, neural networks, gaze detection, text tagging}

\input{0_abstract_and_teaser}


\end{frontmatter}

\begin{figure*}
\centering
\includegraphics[width=\linewidth]{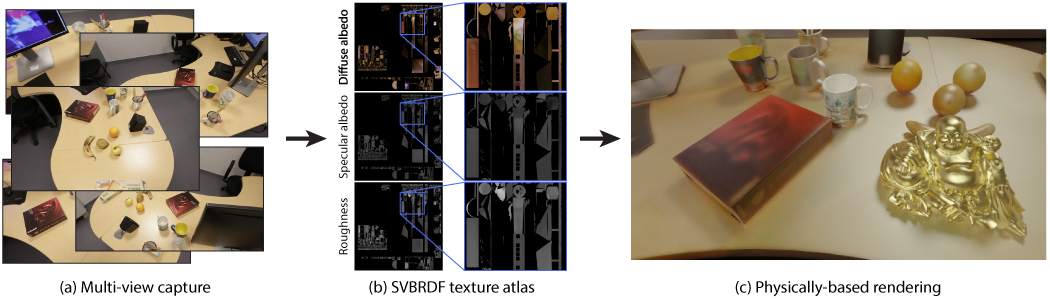}
\caption{
We present a method which takes multiple photographs of a scene
% and a retopologized mesh 
%\TODO{ADD THE MESH BELOW IMAGES IN (a)} 
%SP{We said we won't introduce the mesh here but in separate figures so I'm removing mesh reference here}
as input (a) and predicts surface materials in the form of material maps corresponding to each input view. 
Merging these image-space predictions in texture space yields a texture atlas (b) that can be mapped onto retopologized geometry to produce \emph{digital 3D assets} ready for full physically-based rendering, i.e., rendering from any viewpoint, changing/adding lights and objects, for example in (c) the golden statuette and the white mug have been added to the scene that has modified lighting and is rendered from a viewpoint not in the input. 
%The SVBRDF maps are multi-view consistent and can be pooled in texture space to obtain scene-scale SVBRDF parameters. Using the obtained maps, we show applications such as relighting, object insertion and material manipulations enabling realistic scene augmentations.}
\label{fig:teaser}}
%\vspace{0.5cm}
\end{figure*}

%\linenumbers
\input{1_intro}

\input{2_related_work}

\input{3_overview}

\input{4_method}

\input{5_evaluations}

\input{6_conclusion}

\input{acknowledgement}

%\section*{Acknowledgments}
%Funding - ERC, Cluster[https://wiki.inria.fr/ClustersSophia/Usage_policy#Acknowledgment] - NEF
%The research was funded by the ERC Advanced grant FUNGRAPH No 788065 (\url{https://project.inria.fr/fungraph/}).
%The authors are grateful to Inria Sophia Antipolis - M\'editerran\'ee ``Nef" computation cluster and the OPAL infrastructure from Universit\'e C\^ote d'Azur for providing resources and support.

%\input{appendix}

\newpage
%\bibliographystyle{ACM-Reference-Format}
%\bibliography{sample-bibliography}
%\bibliographystyle{cag-num-names}
%\bibliography{deep_materials_cag} 

\input{deep_materials_cag_arxiv.bbl}
\end{document}

%% file: 0_abstract_and_teaser.tex
\begin{abstract}
%  A clear and well-documented \LaTeX\ document is presented as an
%  article formatted for publication by ACM in a conference proceedings
%  or journal publication. Based on the ``acmart'' document class, this
%  article presents and explains many of the common variations, as well
%  as many of the formatting elements an author may use in the
%  preparation of the documentation of their work.
The movie and video game industries have adopted photogrammetry as a way to create digital 3D assets from multiple photographs of a real-world scene.
But photogrammetry algorithms typically output an RGB texture atlas of the scene that only serves as visual guidance for skilled artists to create material maps suitable for physically-based rendering.
%\SP{\sout{SVBRDF} (they serve as visual guidance for more than just SVBRDF maps, such as height map, AO map, irradiance maps etc.)} \NEW{feature / property} maps suitable for physically-based rendering \NEW{often using crude hand-crafted heuristics}. 
We present a learning-based approach that automatically produces digital assets ready for physically-based rendering, by estimating approximate material maps from multi-view captures of indoor scenes that are used with retopologized geometry.
We base our approach on a material estimation Convolutional Neural Network (CNN) that we execute on each input image. We leverage the view-dependent visual cues provided by the multiple observations of the scene by gathering, for each pixel of a given image, the color of the corresponding point in other images. This image-space CNN provides us with an ensemble of predictions, which we merge in texture space as the last step of our approach. 
Our results demonstrate that the recovered assets can be directly used for physically-based rendering and editing of real indoor scenes from any viewpoint and novel lighting. Our method generates approximate material maps 
%is fully automatic and can obtain the SVBRDF atlases 
in a fraction of time compared to the closest previous solutions.

\end{abstract}

%% file: 1_intro.tex
\section{Introduction}

%\begin{itemize}
%    \item Capturing assets, Retopology; Materials are hard and time-consuming to design. Material estimation focused on patches or objects (too much?), or costly optimization.
%    \item We present a learning based method to rapidly estimate materials at scene-scale, without expensive optimization
%\end{itemize}

While physically-based rendering is now a mature technology~\cite{fascione17}, creating the digital assets to be rendered remains a major bottleneck in the creative industry.
%Creating digital assets is one of the most challenging tasks in Computer Graphics. 
Photogrammetry has gained popularity to create digital assets from real-world scenes. 
A popular workflow consists in first capturing multiple photographs of the scene,
then using multi-view stereo algorithms to compute an approximate 3D model from these photographs.
The approximate geometry is then manually edited to create models compatible with traditional rendering pipelines -- a task known as \emph{retopology}~ \cite{unity2017photo}.
%\SP{It's unclear whether retopology only implies cleaning up geometry and UV-unwrapping or it also includes creating material texture atlases which are included in the term ``assets" as well. It's the former}. 
%\SP{I'd say MVS is for geometry only. I'll rather use photogrammetry instead of MVS.}
Unfortunately, existing photogrammetry solutions typically output a simple RGB texture of the scene with baked-in lighting, which only serves as a crude initialization for artists who need create rich materials maps used by downstream physically-based renderers. 
Creating these maps involves significant manual work, including removing shading, shadows and highlights to form the diffuse albedo, and guessing specular strength and roughness parameters over different surfaces.
%by manually removing shading, shadows and highlights to obtain a diffuse albedo, and who guess specular strength and roughness parameters that form the material maps used by downstream physically-based renderers.
We propose a learning-based approach that addresses this difficult task by augmenting the photogrammetry workflow by \emph{automatically} estimating approximate material maps from multiple photographs of an indoor scene. Our goal is to provide assets that directly allow plausible renderings of the captured scene. Specifically, the output of our method are approximate Spatially-Varying Bidirectional Reflectance Distribution Function (SVBRDF) \emph{texture atlases} which, combined with retopologized geometry, forms a digital asset ready for physically-based rendering of indoor scenes. We call these \emph{material maps} from now on.

%since it creates a first approximation of the scene, followed by \emph{retopology} where artists create a clean "renderer-friendly" version of geometry~\cite{Unity}. A complex and time-consuming step is then required to manually design and apply materials to the scene. 

Learning-based methods for material estimation have focused on pictures of flat surface patches~\cite{deschaintre2018,guo2021} or on \emph{single images} of isolated objects \cite{li2018learning} and scenes \cite{li2020invrender}, for which the prediction can be efficiently performed in image-space using convolutional neural networks (CNNs). 
In contrast, approaches based on inverse rendering compute accurate material parameters in object or texture space \cite{yu1999,merlin2021} to benefit from observations from multiple viewpoints. But the underlying optimization is expensive and needs to be recomputed for every new scene.
We present the first method that combines
%\sout{Our method demonstrates the benefit of combining} 
ideas from these two streams of research. 
On the one hand, we leverage the strength of image-space CNNs to predict approximate material parameters for each photograph of the scene. On the other hand, we exploit multi-view information by gathering, for each pixel in a photograph, observations of the same scene point in other photographs. Furthermore, we aggregate the predictions given by each photograph into a common texture space to form the final texture atlas. Our method thus offers the speed of learning-based material estimation previously applied for single images, for the much harder scene-scale material estimation problem.

Our method addresses
%\sout{practical solutions to} 
several difficulties raised by the long-standing challenge of scene-scale material estimation.
First, in contrast to single-image methods~\cite{deschaintre2018}\cite{li2018learning}\cite{li2020invrender}, our multi-view setting receives a varying number of observations per pixel to be processed by the CNN. We overcome this difficulty by computing a fixed number of color statistics, which forms the initial feature maps that we feed to the CNN.
Second, photographs of indoor scenes exhibit complex interactions between geometry, lighting and materials via indirect illumination. 
Prior work on inverse rendering model these interactions using approximate global illumination (GI) to jointly recover shape, materials and light~\cite{li2020invrender}. 
In contrast, we consider a scenario where geometry is reconstructed with photogrammetry and retopology, such that we only need to recover material appearance. 
%\NEW{Rather than approaching this as an unsupervised inverse rendering problem where global illumination must be estimated accurately to match the input observations, we train a neural network to predict materials maps of similar appearance to the inputs using ground-truth maps for supervision. To compare the predictions to the ground truth materials, we use a rendering loss which only operates via local camera-space relighting, avoiding the underconstrained estimation and expensive computation of GI altogether.}
Rather than approaching this as an inverse rendering problem where global illumination must be estimated accurately to match the input observations, we train an illumination agnostic network to produce materials maps of similar appearance to the inputs. To compare the predictions to the ground truth materials, we use a rendering loss which only operates via local camera-space relighting, avoiding the underconstrained estimation and expensive computation of GI altogether.
%Under these assumptions, a \emph{local} lighting model is sufficient to evaluate whether the predicted materials have a similar appearance to ground-truth material maps \NEW{instead of computing the full GI to reproduce the input image.
%Using a local lighting model to compute renderings for our rendering loss during training} enables our network to be agnostic of input lighting conditions, and thus our method does not require knowledge about scene illumination for material maps estimation.
%\SP{I feel here it is unclear that we use a rendering loss to train our network and the local lighting model is used to compute the renderings in the loss. We do not need to emphasize it as novel but we do need to make it clear.}
The third challenge is building a synthetic training dataset suitable for scene-scale approximate material estimation; we created a dataset from professionally modeled scenes, and provide a framework that allows the generation of new datasets for this task.

In summary, our contributions are:
\begin{itemize}[topsep=0pt]
    \item A deep neural network architecture for material estimation that exploits scene-scale multi-view input. %\item A rendering loss that includes area lighting with spherical Gaussians to better capture specular information. 
    \item A proof-of-concept solution allowing fast, scene-scale material estimation to produce digital assets suitable for physically based rendering and editing of real indoor scenes, that integrates seamlessly into the current photogrammetry workflow.
    \item A scene-scale synthetic dataset with ground truth SVBRDF maps and the tools to generate it, used to train our multi-view material estimation network.
\end{itemize}

\noindent
We evaluate and illustrate our method on synthetic scenes that allow quantitative analysis of our algorithmic choices, and show first results on captured real scenes. We demonstrate that our automatically estimated material map atlases -- albeit approximate -- are of sufficient quality to allow physically-based rendering of the captured scene with novel lighting conditions and scene editing (see Fig.~\ref{fig:teaser}(c),~\ref{fig:results2}). 
%\GD{or scene if need be}
We will provide the source code to our system, including all the tools required to generate the training dataset from commercially-available models.

%% file: 2_related_work.tex
\section{Related Work}
\label{sec:related}

We discuss the two domains that inspired our approach -- optimization-based and learning-based methods for practical material estimation.
%On the one hand, learning-based methods for material capture often focus on planar surface patches or isolated objects  captured under flash lighting. On the other hand, methods to recover materials over entire indoor scenes rely on expensive optimization and inverse rendering. We seek to bridge these two domains by offering a learning-based solution that can replace optimization methods, or that can serve as a high-quality initialization.
We refer the interested reader to surveys on material capture \cite{garces2022surveyintrinsic, guarnera2016} and inverse rendering \cite{tewari2022advances, kato2020differentiable} for more general discussions of these broad topics.

\subsection{Optimization-based material capture}
The recent development of differentiable rendering algorithms has reinvigorated research on inverse rendering, as pioneered by Yu et al.~\cite{yu1999} for scene-scale material recovery. 
Given a 3D model of the scene, modern approaches estimate material and lighting by simulating complex global illumination effects using differentiable path-tracing~\cite{azinovic2019, merlin2021, haefner2021}. We see our approach as complementary to such optimization-based algorithms that work on scene-scale. On the one hand optimization-based algorithms are capable of recovering more precise information by minimizing the difference between the input images and the images re-rendered from the estimated materials. On the other hand, such a minimization typically takes $10$-$12$ hours to converge due to complex global illumination computations and is highly sensitive to initialization.
%, which our learning-based approach can provide in a fraction of the time given similar input. 
Our approach could speed-up these optimization methods by providing an initialization that is much closer to the end result compared to the random material maps that are typically used.
Methods that also optimize for geometry have so far been limited to convex isolated objects often with specific lighting/capture constraints~\cite{wu2016, nam2018, goel2020, ma2021, luan2021, bi2020deep}. In contrast, we take multiple unconstrained sparse viewpoints of the scene resulting in a variable number of observations for different scene regions and complex visibility issues due to inexact geometry. 

\subsection{Learning-based material capture}

Recovering material appearance from a few observations is an ill-posed problem for which machine learning offers practical solutions. By leveraging large datasets of images paired with ground truth material maps, deep convolutional networks can be trained to predict per-pixel material parameters given a single picture of a flat surface patch \cite{li2017modeling,li18mfm,deschaintre2018,gao2019, guo2021,zhou2021, henzler2021neuralmaterial}. Such methods were later extended to predict material, depth and normal maps of isolated objects \cite{li2018learning, boss2020} and even of indoor scenes \cite{li2020invrender} from a single image. Ours is the first method to take wide baseline, scene-scale multi-view input under unknown indoor lighting for material estimation. Most related to our goal is the concurrent work by Li et al.~\cite{Zhengqin22}, who take as input a \emph{single image} of an indoor scene along with a 3D model of that scene, and build upon single view material prediction \cite{li2020invrender} to assign procedural material models to object parts. Our work is complementary, as we explore material prediction in indoor scenes under a multi-view capture scenario.

Our key insight is that the multiple images that are typically captured for photogrammetry offer complementary observations of material appearance. However, such multi-view information needs to be properly aggregated to be fed to a neural network, and to be assembled to form a valid texture atlas for later use in rendering engines. 
Prior methods on multi-image material prediction only partially address these challenges since they focus on small planar patches and assume that each point is visible in all input images \cite{deschaintre2019,guo2020,asselin2020}, which is not the case when dealing with complex scenes where parts are frequently occluded or out of the field of view of many of the input photographs. While Ye et al.~\cite{ye2021} describe how to combine image warping and max pooling to handle videos captured with a moving camera, they again focus on planar surfaces free of occlusion, and recover material parameters over a reference frame rather than over all input views of a 3D scene. We propose a solution based on image re-projection and pixel statistics to process multi-view inputs with a standard CNN architecture, and to merge multiple predictions into a single texture atlas. 
%\SP{Tobias mentioned this Kavita Bala's 2013 paper~\cite{photometricAO} to cite as a use case of using color statistics  as well.}
%\GD{Not completely sure about this @adrien ?}
%\AB{Right, but we don't use the same statistics as theirs (they look at variance under varying lighting).}

Neural representations recently emerged as an effective solution to relight 3D content captured from multiple photographs \cite{nerv2021,zhang2021nerfactor,zhang2021physg, boss2021nerd, boss2021neuralpil}.
However, these novel representations are not compatible with the well-established photogrammetry workflow, where artists seek to create triangular meshes and texture atlases compatible with downstream industry-standard rendering engines. While Philip et al.~\cite{philip2021} also feeds color statistics as one of the many multi-view information to a neural renderer to perform novel-view synthesis with relighting, we predict explicit material parameters in the form of material texture maps and we assemble the predictions given by multiple views in a common texture space which is readily available to the user for further editing as desired. This post processing flexibility is missing from prior works. More recently, Munkberg et al.~ \cite{munkberg2021nvdiffrec} combine neural and traditional representations within a differentiable rendering framework to recover a triangle mesh, an SVBRDF texture and an environment map, but their approach has only been demonstrated on isolated objects. 

Training methods like ours require large amounts of photorealistic images with ground truth material map labels. Such a dataset is infeasible to capture so we rendered visually realistic synthetic scenes with variations in lighting, materials, geometry and viewpoints. While several datasets of indoor scenes have been described, many only provide images rendered from pre-defined viewpoints and do not allow the generation of new images, as is the case for \emph{OpenRooms}~\cite{openrooms2021}. Other datasets do not include the labels we are interested in, such as \emph{Hypersim}~\cite{hypersim2020} that provides diffuse albedo maps and a non-diffuse residual term, which is not directly compatible with existing BRDF models suitable for physically-based renderers.
%but no specular maps \SP{HyperSim provides a ``non-diffuse residual term" which can be taken as specular highlights. Maybe a good place to point out how different datasets use different material models (which is the case with OpenRooms as well) and to gain better control on the data/buffers we generate for training we created our own dataset. Related point against OpenRooms and Hypersim both is that they do not provide specular maps, due to the different models they use}.
%\GD{These previous datasets use different lighting models which are not always compatible with PBR and thus are not suitable for our purposes.}
We built a dataset tailored to multi-view material estimation in indoor scenes, by developing an asset generation system that assembles objects from synthetic scenes modeled in Autodesk 3DS Max and then rendering with Mitsuba~\cite{jakob2010mitsuba}. We hope our dataset generation tools will help foster research on scene-scale material estimation and other scene-scale learning-based tasks.

%% file: 3_overview.tex
\begin{figure*}[!t]
\includegraphics[width=\linewidth]{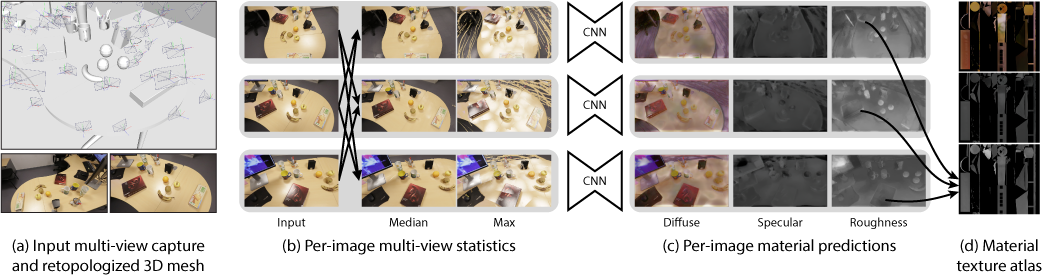}
\caption{\label{fig:overview}
Overview of our method. We take a multi-view dataset, calibrated cameras and a retopologized mesh as input (a). 
We re-project information from multiple views into each input view to compute color statistics for each observed point (b).
We feed this multi-view information into a convolutional neural network to predict material maps for each view (c).
We finally merge this ensemble of inferred maps into texture space (d) to form a scene-scale material texture atlas.
%Our deep learning method then infers SVBRDF maps for each input image (b), using re-projected information from multiple views, and a bilateral solver for specular and roughness maps. We train this network using a dataset of synthetic images of scenes with a variety of geometries and materials (c). We then combine these inferred maps into texture space (d), resulting in a fully renderable digital version of the scene, that can be rendered freely under new lighting or with additional assets (e).
}
\end{figure*}

\section{Overview}
\label{sec:overview}

Our method takes as input multiple casually-captured unstructured wide-baseline images of a scene from different viewpoints using a DSLR camera. 
This results in multi-view observations, but in some cases capture is incomplete: In particular, lighting and some parts of the scene might be unobserved. Similar to commercial photogrammetry pipelines~\cite{unity2017photo}, first we obtain camera calibration and a rough multi-view stereo mesh using RealityCapture~\cite{reality2016capture}, followed by re-topology where an artist creates a clean version of the mesh, suitable for rendering (Fig.~\ref{fig:overview}(a) and~\ref{fig:geometry}). This reliable 3D geometry is used as input to our method.
%As a result we obtain smooth mesh with holes filled and sharp occlusion edges. 
%We first describe the material model used in our method followed by an overview of the pipeline and the training dataset.
%Then we give an overview of our potential contributions for predicting SVBRDF maps given multi-view information as input to the network in Section.
Given the multiple input images and the corresponding geometry, our goal is to produce an atlas of material parameters, i.e., spatially-varying diffuse albedo~$D$, specular albedo~$S$, and roughness~$R$ for a Cook-Torrance BRDF model~\cite{ct82}. We do not estimate normal maps since we focus on indoor scenes composed of large surfaces seen at a distance, for which our retopologized geometry provides sufficiently accurate normals (Fig.~\ref{fig:overview}(a) top and~\ref{fig:geometry}).

We achieve this material estimation task in two main steps, illustrated in Fig.~\ref{fig:overview} (b - d).
The first step relies on an image-space CNN to predict material maps for each input image separately. For each view, we use the 3D geometry to reproject image colors from other views and deduce color statistics (minimum, median and maximum color) that summarize the view-dependent appearance of each pixel. We complement these statistics with geometry buffers (surface normals and depth). In practice, we found it beneficial to split the prediction task into two tracks, one responsible for the prediction of diffuse albedo and one responsible for the prediction of specular albedo and roughness. 
%\AB{I suggest to add motivation for these choices later on. Same for the description of progressive training and bilateral solver.}

%We first train separate tracks for roughness/specular maps, and then for diffuse albedo \SP{INCORRECT. 
%We first train the diffuse albedo track to get a good baseline for underlying surface color/texture and then train the roughness/specular maps track, followed by a joint fine-tuning step (Fig.~\ref{fig:overview}(b). 

%Finally, the roughness and specular layers are processed with a learned bilateral solver. We provide details of the layers and the network architecture in Sec.~\ref{sec:network}, which is trained on our synthetic scene dataset (c) \SP{Not shown in current overview}, described in Sec.~\ref{sec:dataset}. 
The second step of our method aggregates the per-view predictions into a common texture space to form the final atlas. We use simple median filtering to select a consensus from the ensemble of predictions given by all views where a surface point appears. Mapping this atlas onto the retopologized mesh gives a complete asset that is compatible with traditional physically-based rendering (Fig.~\ref{fig:teaser}(c), \ref{fig:results2})\CK{, including full editing capabilities such as changing the lighting and inserting new objects}.

%% file: 4_method.tex
\section{Multi-View Aware Deep Material Estimation}
\label{sec:network}

Our problem is estimating \emph{scene-scale} material properties from a \emph{multi-view dataset} under \emph{unknown lighting}, as opposed to the several successful deep learning methods for estimating SVBRDF maps: These start from one or a few images~\cite{li2017modeling,li18mfm,deschaintre2018,guo2021,zhou2021,deschaintre2019,asselin2020}, typically for small planar patches of materials lit by a flash. 
We tackle the scene-scale material estimation problem by first processing each view with a CNN similar to the one used by Deschaintre et al.~\cite{deschaintre2018, deschaintre2019}; We explain how we adapted this architecture to our use-case in Sec.~\ref{sec:architecture}.

\begin{figure}[t!]
\centering
\includegraphics[width=\linewidth]{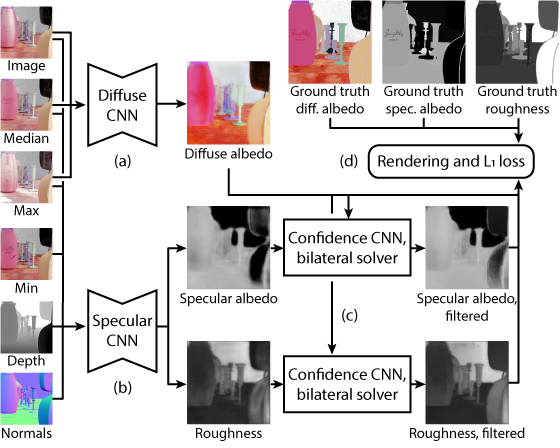}
\caption{
    \label{fig:architecture}
    Overview of our deep learning architecture and training procedure.
    We split the material estimation task into two tracks, one for the diffuse albedo and the other one for specular albedo and roughness (a, b). In addition, we filter the specular albedo and roughness with a bilateral solver (c). Finally, we compare the predicted maps with ground truth maps using a $\mathcal{L}_{1}$ loss as well as with a rendering loss that assesses the appearance of the materials under several lighting and viewing conditions (d). 
}
\end{figure}

%, trained on a synthetic dataset of realistic images paired with ground-truth diffuse albedo, specular albedo and specular roughness maps. 
The much harder scene-scale problem precludes the use of co-located flash lighting;
Since we cannot benefit from the rich visual cues given by this mode of capture, our originality is to instead leverage visual cues provided by multi-view observations. 
Figure~\ref{fig:architecture} illustrates the main components of our architecture and its training procedure.
%The basic building block of our approach is inspired from Deschaintre et al. \cite{deschaintre2018}, which uses a U-net architecture augmented with a global track, allowing inference of albedo, roughness, specular and normal maps from an image, trained on a synthetic dataset. 

\subsection{Network Inputs}
While we run our CNN on each input view separately, we feed it with multi-view information obtained by re-projecting a set of neighboring views to the current view.
For each pixel, we select at most 12 views where the corresponding point is visible, and for which the view direction is most closely aligned to the surface normal, as those views are less prone to grazing angle observations. We also add a distance term to favor cameras nearer to the surface. 

\begin{figure}[!t]
\center
\includegraphics[width=0.8\linewidth]{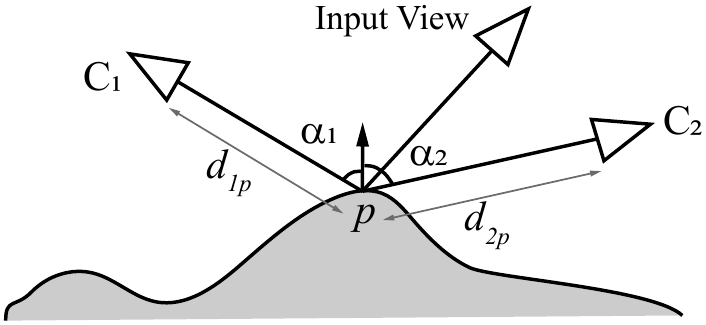}
\caption{
Camera selection for each surface point $p$. A cost function is used to select the 12 best views that consists of a visibility term and a distance term which favors normal-aligned views and cameras close to the surface. }
\label{fig:cost}
%\Description{Few examples of renderings obtained with our synthetic dataset showing realistic geometry, materials and lighting conditions along with the variations.}
\end{figure}

We rank and pick the top-12 views by minimizing the cost composed of a view term and a distance term as shown in Fig.~\ref{fig:cost}, specifically for an observation at $p$:
\begin{equation}
\mathrm{cost}_i ~=~ cos~ \alpha_i + \frac{d_{ip}}{max(d_{jp})} \quad \quad \forall j \in \{1, N\}
\end{equation}
\noindent
where $d_{ip}$ is the distance of point $p$ to camera $i$, $N$ is the total number of cameras, and $\alpha_i$ is the angle between the normal at surface point $p$ and view direction of camera $i$.

Since each pixel might receive a different number of observations as not all surface may be visible in 12 or more views (this is frequently observed in the case of background pixels), we summarize this multi-view information as a fixed set of images corresponding to simple color statistics, i.e. median, maximum and minimum. 
%\SP{Reviewers mentioned to formally define the statistics, which can be done here if required. I'm not convinced we need it though.}

The maps that form a material map atlas have different characteristics, and can be inferred from different visual cues. On the one hand, the diffuse albedo map needs to maintain sharp texture features while being free of view-dependent highlights, shadows and indirect light. On the other hand, the specular albedo and roughness values are often nearly constant over parts of objects made of the same material, and are conveyed by highlight information from different views. These observations motivate us to predict the diffuse and specular parameters via two different tracks, and to feed each track with different visual information.

For the diffuse albedo, we complement the input image with the median image, as the median rejects highlights that only appear in a few images. We also include the maximum image to help the network locate shiny areas where highlights might need to be suppressed even if present in many images. 

In addition to the images used by the diffuse track, we feed the specular track with the minimum image as it further helps locate parts where the color changes significantly across views. We also provide the specular track with normal and depth maps, which delineate different objects that often have different specular values. Our experiments revealed that providing all these extra images to the diffuse track degrades its prediction, as the network struggles to select the relevant image structures among too many visual channels. In addition, reprojection errors due to the approximations in retopologized geometry can be problematic (see Sec.~\ref{sec:results}).

Finally, we favor smooth specular maps by post-processing their predictions with a differentiable bilateral solver \cite{barron2016BS, li2020invrender} guided by the predicted diffuse albedo. 
This edge-aware smoothing attenuates discontinuities due to reprojection misalignments.
Unlike Li et al.~\cite{li2020invrender} we do not run our diffuse albedo maps through the solver as we observed this leads to over-smoothing of diffuse albedo maps resulting in loss of texture details.

\subsection{Network Architecture}
\label{sec:architecture}
Our network architecture is based on the ones by Deschaintre et al.~\cite{deschaintre2018, deschaintre2019} which were also designed for material estimation but works on a single flat patch of size 256x256x3. Their architectures follow the widely popular U-Net encoder-decoder~\cite{ronneberger2015unet}, to which a fully-connected track responsible for processing and transmitting global information was added. 
We maintain the same U-Net architecture augmented with the global track but we half the number of feature layers from 8 to 4 to make the networks lighter with the feature counts in the encoder downscaling layers of 64, 128, 256, and 512. We did not observe any significant degradation in the maps due to this reduction in network capacity.
We follow the same downsampling and upsampling process as described in~\cite{deschaintre2018} with the feature counts used in reverse order for the decoder. 
Instance normalization is used for stability and we also regularize by applying dropout at 50\% probability on the last three layers of the decoder.

The main difference lies in the input/output; Specifically 
our network architecture differs in two ways. 
First, we separate out the network into two tracks one each for diffuse and specular maps. 
The two tracks serve different purposes with the diffuse track predicting only diffuse albedo maps, while the specular track outputs the specular albedo and roughness maps. 
The two tracks are two separate networks. 
While training, we can either train one of the networks without sharing any information between them or both networks by jointly computing the loss on output of both networks.
We use individual training to train the networks and joint training to fine-tune the networks post-training. 
Please see Sec.~\ref{sec:implementation} for training details. 
Second, since we feed the network with multi-view information in the form of image statistics, we have more input channels compared to $3$ channels for the previous networks. Concretely, since we feed the median, maximum and input image to the diffuse track which thus has $9$ input channels, while for the specular track, we also provide the minimum image, depth and normal buffers resulting in $18$ input channels.
%\TODO{UNCLEAR: THERE WE SAY "AS A FIXED SET OF IMAGES", WHICH CONTRADICTS THIS; CLARIFY ?}
%Deschaintre at al.~\cite{deschaintre2019} also provided pixel coordinates as extra channels which is not required by our method, since we have geometry}\TODO{VERIFY}.

For the network used to predict confidence channels to guide the bilateral solver, we follow the same CNN architecture and hyper-parameters for the solver as used in previous works~\cite{li2020invrender, barron2016BS}.
We include detailed breakdown of the network architecture and the parameters used in the supplemental PDF.

\subsection{Loss Function}

Many inverse rendering methods supervise their predictions of geometry, material and lighting by comparing re-rendered images to input images using a differentiable \emph{rendering loss} \cite{li2018learning,boss2020}. At scene-scale, such a rendering loss needs to model global illumination effects present in the input \cite{li2020invrender}. We depart from this family of methods by focusing on a scenario where geometry is given, such that our task boils down to predicting material maps only. In this context, we can supervise our method by comparing our prediction to ground truth SVBRDF maps rather than by attempting to reproduce the input. A local lighting model is sufficient for this purpose, as was originally proposed by Deschaintre et al. \cite{deschaintre2018} in the context of planar surface patches. While Deschaintre et al. use point and directional lights, we improve on their approach by incorporating distant area lights modeled as spherical Gaussians.
%, which provide a broader spectrum of highlight positions and shapes, which in turns yield more accurate predictions of specular parameters (see Sec.~\ref{sec:evaluation}).

Concretely, we use a simple differentiable renderer that takes as input the material maps along with the normals of the geometry. Our goal is to compare the local appearance of our predicted material maps to the local appearance of the ground truth SVBRDFs. 
%\SP{Can we use material parameters/maps instead of SVBRDFs? It's confusing since researchers expect SVBRDFs to include normal maps as well. Nimier-David et al.\cite{merlin2021} avoids using SVBRDF maps completely too and uses material parameters wherever required which I feel goes a long way in expectation management.} 
To do so, we render the prediction and the ground truth under several viewing and ligthing conditions and compare the resulting images under the $\mathcal{L}_{1}$ norm and E-LPIPS perceptual metric $\mathcal{L}_{E}$~\cite{kettunen2019lpips} after applying a log transform ($I' = log(0.1 + I)$) to compress the dynamic range of the renderings. 
Following Deschaintre et al., we generate random viewing conditions by sampling view vectors over the hemisphere centered around the original view direction from which the input image was rendered. We then generate lighting conditions likely to produce highlights by positioning a point light in the mirror direction of the view vector. Finally, we also create extended light sources by generating a mixture of $5$ Gaussian lights with random width, color and direction distributed over the hemisphere.

We implement the shading of a point under distant area light sources by using the spherical warp introduced by Wang et al.~\cite{wang09}. The light, as well as the BRDF, are approximated as two Spherical Gaussians, for which a fast closed-form convolution exists. Using this approximation allows us to include extended light sources in the rendering loss without losing computational efficiency for training.
Our final rendering loss averages the image differences obtained with three point-wise lighting conditions and with three extended lighting conditions. 

In addition to the rendering loss, we also use $\mathcal{L}_{1}$ and $\mathcal{L}_{E}$ to compare the individual predicted maps to their respective ground-truth. Denoting $I$ a rendered image, $D$ the diffuse albedo, $R$ the roughness, and $S$ the specular albedo, 
%terms on both the rendering $\mathcal{R}$ and the roughness $\mathcal{M_R}$, specular $\mathcal{M_S}$ and albedo $\mathcal{M_A}$ maps. Diffuse and specular maps undergo a log transform ($log(\epsilon=0.1 + I)$) before computing the loss. 
the total loss we use is thus:
\begin{align}
\mathcal{L} & = [\mathcal{L}_{E}(I)+ \mathcal{L}_{1}(I)] \nonumber\\ 
& + \mathcal{L}_{E}(D)  + \mathcal{L}_{E}(R) + \mathcal{L}_{E}(S)  \nonumber\\
& + \lambda (\mathcal{L}_{1}(D) + \mathcal{L}_{1}(R) + \mathcal{L}_{1}(S)) 
\end{align}
where $\lambda$ is $0.1$.

\subsection{Merged Renderable Scene Assets}
\label{sec:texture-space}
The second step of our method merges the material maps predicted over each input view to form a single, object-space material map texture atlas suitable for rendering.
We leverage the retopologized 3D mesh to identify which texel corresponds to each pixel in all input views. We select the final value of each texel as the median value of all its predictions. 
This median filter is especially effective at removing erroneous predictions that are not consistent across views (see Sec.~\ref{sec:analysis})
%(Figure~\ref{fig:map-viz}),  
including the ones due to re-projection artifacts caused by approximate geometry and camera calibration in real-world scenes.

\section{Synthetic Training Dataset}
\label{sec:dataset}

% Dataset
We trained our method by generating a dataset of synthetic renderings with corresponding ground truth material maps. 
%Instead of using CAD models for geometry, as done by previous datasets~\cite{openrooms2021}, 
We use professional artist-modeled assets in Autodesk 3DS Max with high quality V-Ray materials, that help bridge the gap between training data and real re-topologized scenes.
%\AB{I don't fully buy this argument. What is the distinction between a ``CAD model'' and a ``professional artist-modeled asset''? Also, prior work used Substance or similar materials, which I guess is equivalent to V-Ray materials.}. 
We purchased a set of scenes \footnote{From \url{https://evermotion.org/shop/cat/355/all_scenes/0/0}, Volume 1, 8 and 30.} and extracted basic environments and several different objects that we recombined to create new scene configurations, on which we place objects with different materials. 
%\footnote{\GD{DO THIS The list of scenes and the objectsextracted are provided in supplemental.}}. Each scene has a central surface (typically a table)

We augment the initial artist-generated materials with materials from Deschaintre et al.~\cite{deschaintre2018}, hand-picked to correspond well with the underlying geometry and to cover a wide range of everyday indoor materials such as wood, metal, plastic, rubber, leather, etc.
Each choice of materials and objects provides a scene configuration, for a total of $160$ scene configurations, created from $5$ ``base scenes'' with a set of random object placements. We place area and point lights in the scene, as well as environment maps that typically illuminate the scenes through a window. 
%See the supplementary material for additional details on how we generate material and lighting variations~\GD{VERIF WE DO THIS}. 

\input{figures/synthetic}
We rendered each image using a Cook-Torrance BRDF model with a Beckmann normal distribution, which we have implemented in the Mitsuba physically-based path tracer for full global illumination~\cite{jakob2010mitsuba}.  
We subsequently denoised each rendering using the Optix denoiser~\cite{parker2010optix}. For each image, we also generate the ground truth SVBRDF maps, i.e., diffuse albedo, roughness and specular albedo rendered as images. 
Finally, we pre-compute the per-pixel re-projected color statistics (minimum, median and maximum) for each image in our dataset.
%\GD{VERIFY}

We render each scene configuration under $40$ different viewpoints, yielding a total of $6400$ images at resolution $640 \times 384$ pixels (see Fig.~\ref{fig:dataset} for a small selection).
At training time, we extract random crops of $256 \times 256$ pixels from each image to be fed to the network, which effectively augments the size of the dataset, to around 45,000 individual crops.

%% file: figures/synthetic.tex
\begin{figure}[!t]
\includegraphics[width=\linewidth]{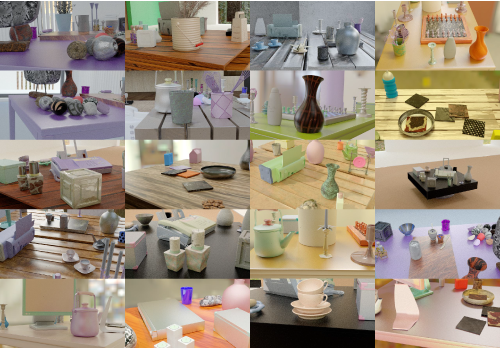}
\caption{
Renderings from our synthetic dataset used for training our network. 
The dataset has a variety of different lightings, materials and viewpoints.}
\label{fig:dataset}
%\Description{Few examples of renderings obtained with our synthetic dataset showing realistic geometry, materials and lighting conditions along with the variations.}
\end{figure}

%% file: 5_evaluations.tex
\begin{figure}[!t]
\includegraphics[width=\linewidth]{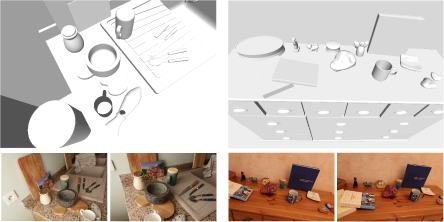}
\caption{\label{fig:geometry}
Retopologized geometry with selected input images for \textsc{real kitchen}~(left) and \textsc{real hallway}~(right) scenes. 
}
\end{figure}

\begin{figure*}[!h]
\setlength{\tabcolsep}{1pt}
\begin{tabular}{cccccc}
%\hline
%\hline
%\multicolumn{6}{c}{Synthetic Veach Ajar} \\
%\hline

 & \textbf{Synthetic Veach Ajar} & & Diffuse & Roughness & Specular \\
{\rotatebox[origin=lB]{90}{\hspace{0.2in} \small{Input Image}}} & 
\includegraphics[width=.235\linewidth]{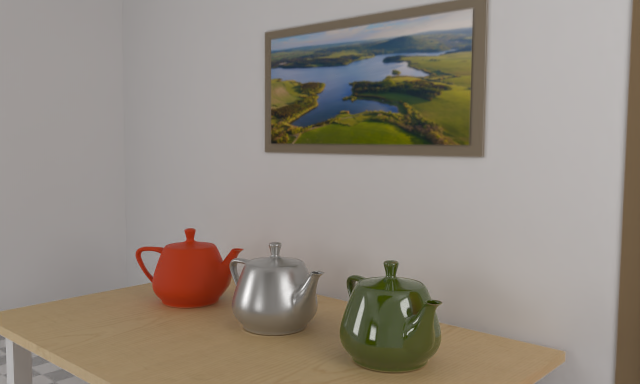} & 
{\rotatebox[origin=lB]{90}{\hspace{0.2in}\small{Ground Truth}}} &
\includegraphics[width=.235\linewidth]{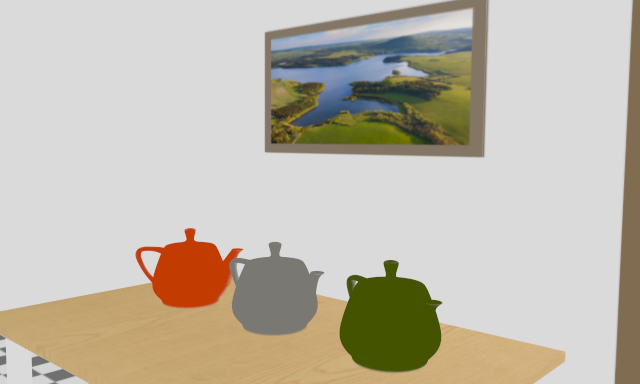} & 
\includegraphics[width=.235\linewidth]{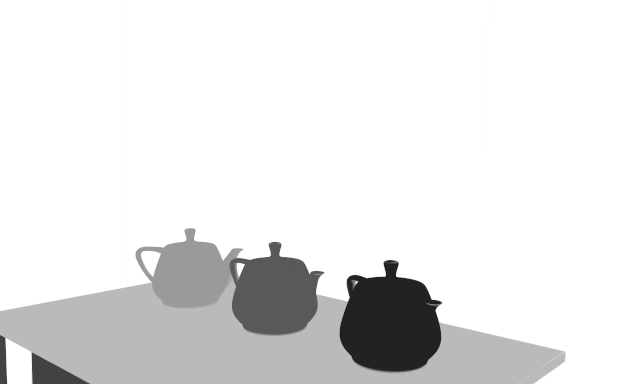} & 
\includegraphics[width=.235\linewidth]{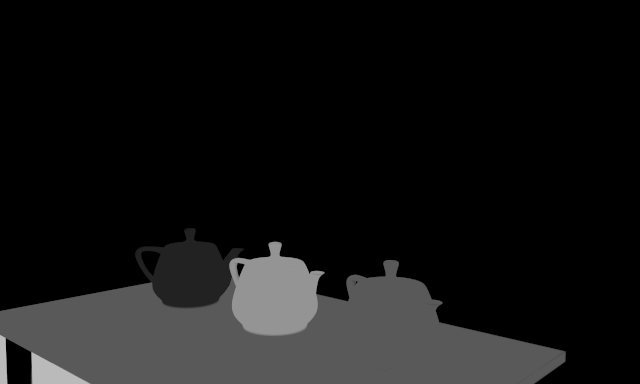} \\

{\rotatebox[origin=lB]{90}{\hspace{0.2in}\small{Re-render}}} & 
\includegraphics[width=.235\linewidth]{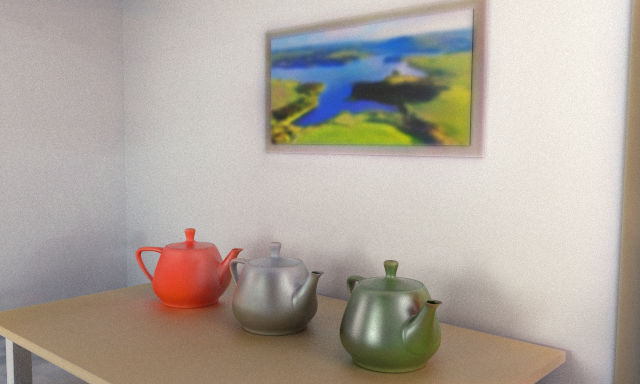} &
{\rotatebox[origin=lB]{90}{\hspace{0.4in}\small{Ours}}} & 
\includegraphics[width=.235\linewidth]{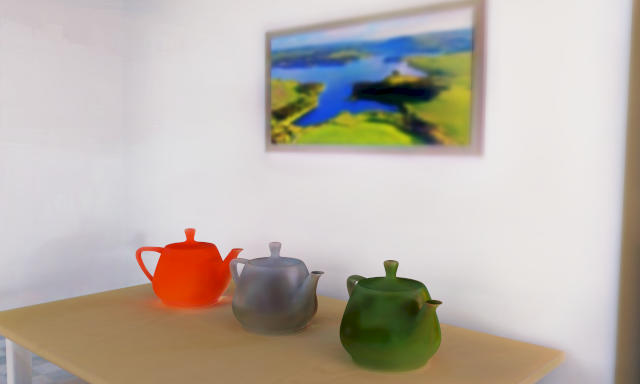} & 
\includegraphics[width=.235\linewidth]{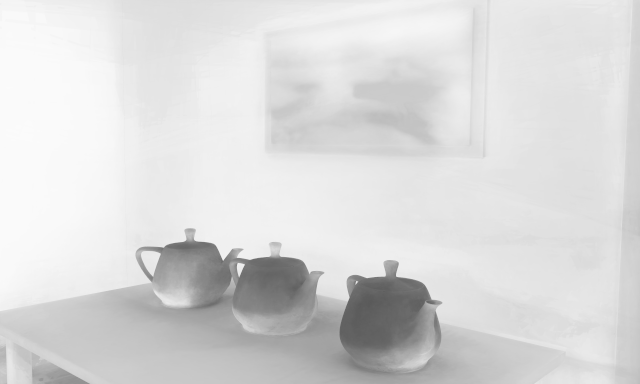} & 
\includegraphics[width=.235\linewidth]{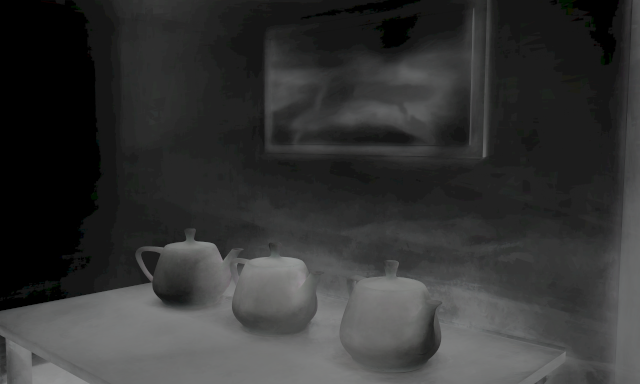} \\

%\hline
%\multicolumn{6}{c}{Synthetic Dining Room} \\
%\hline

 & \textbf{Synthetic Dining Room} & & Diffuse & Roughness & Specular \\
{\rotatebox[origin=lB]{90}{\hspace{0.2in} \small{Input Image}}} & 
\includegraphics[width=.235\linewidth]{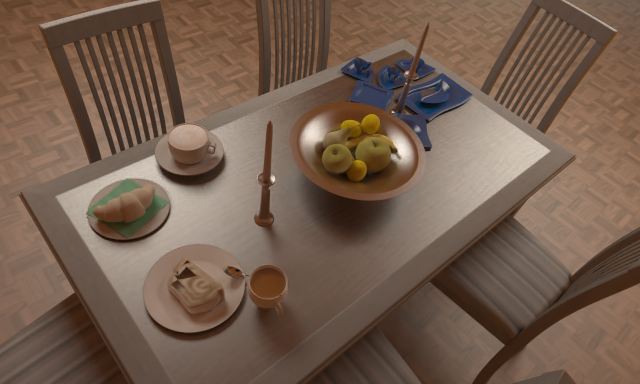} &
{\rotatebox[origin=lB]{90}{\hspace{0.2in}\small{Ground Truth}}} &
\includegraphics[width=.235\linewidth]{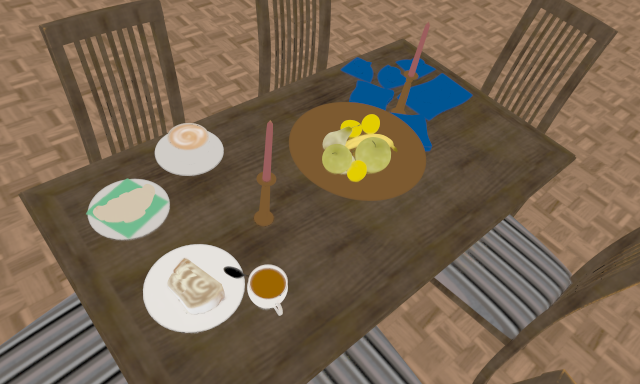} & 
\includegraphics[width=.235\linewidth]{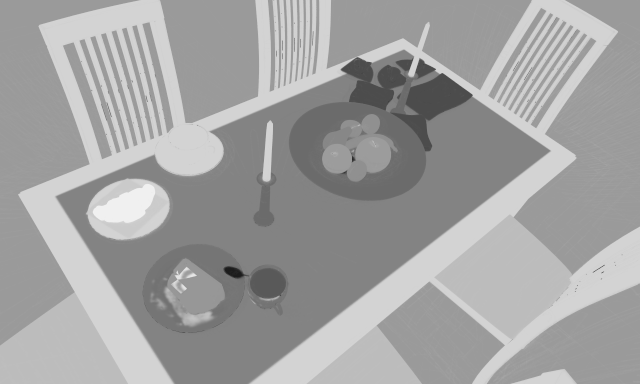} & 
\includegraphics[width=.235\linewidth]{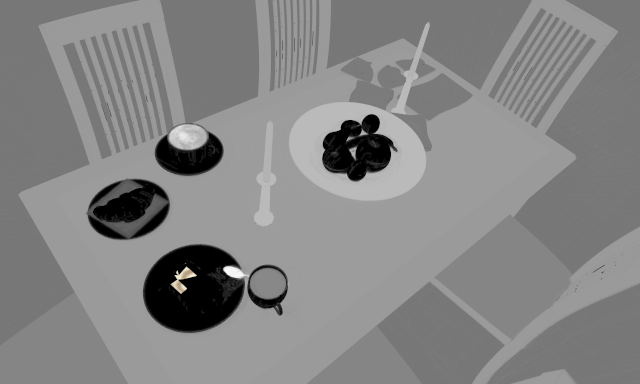} \\

{\rotatebox[origin=lB]{90}{\hspace{0.2in}\small{Re-render}}} & 
\includegraphics[width=.235\linewidth]{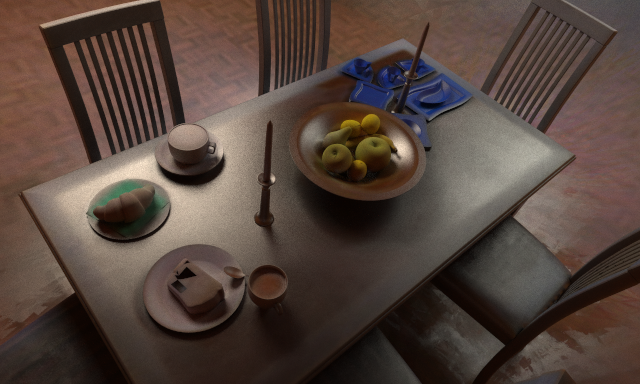} &
{\rotatebox[origin=lB]{90}{\hspace{0.4in}\small{Ours}}} & 
\includegraphics[width=.235\linewidth]{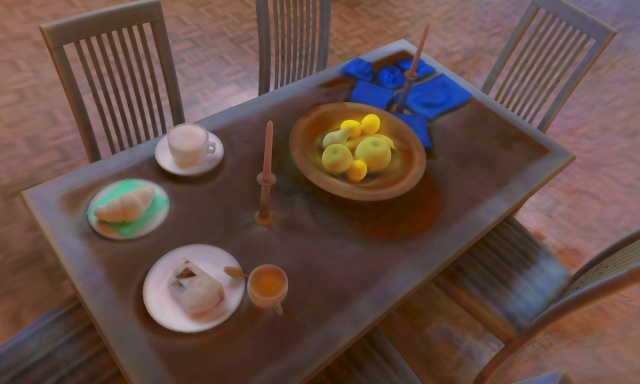} & 
\includegraphics[width=.235\linewidth]{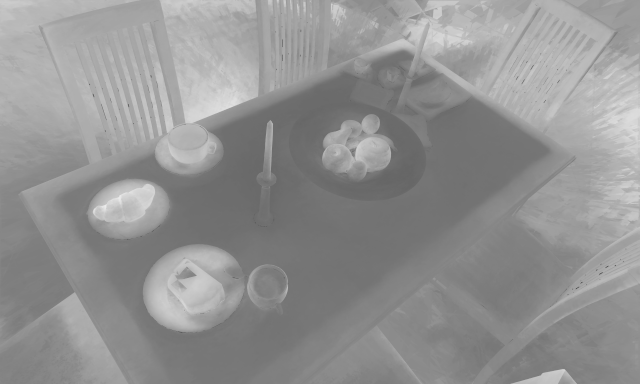} & 
\includegraphics[width=.235\linewidth]{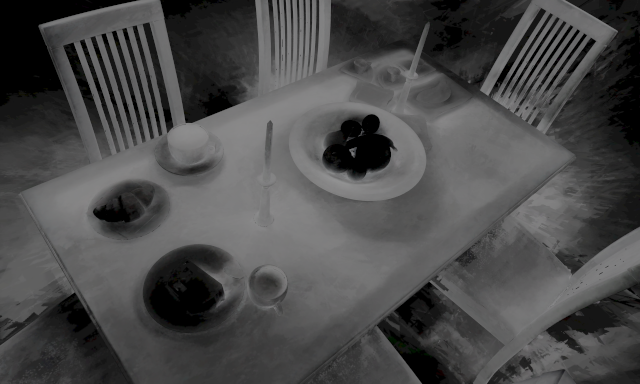} \\

%\hline
\end{tabular}

\caption{
\label{fig:results1}
Results on two synthetic scenes where the ground truth is available. 
For each scene, in the first row we show an input view and the ground truth diffuse, roughness and specular maps for that view. 
The second row shows the re-rendering followed by the maps obtained using our method for the input view shown.
We are able to reproduce renderings which are close to the input view using the approximate material maps generated by our method.
}

\end{figure*}

\begin{figure*}[h!]
\setlength{\tabcolsep}{1pt}
\begin{tabular}{cccc}
%\hline
  & \textbf{Real Office} & \textbf{Real Kitchen} & \textbf{Real Hallway}  \\
%\hline
{\rotatebox[origin=lB]{90}{\hspace{0.5in}\small{Input Image}}} & 
\includegraphics[width=.352\linewidth]{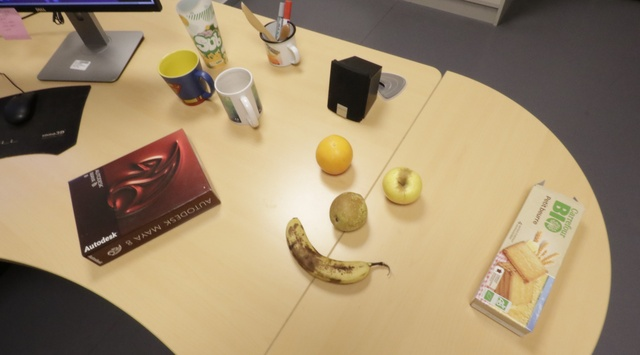} & 
\includegraphics[width=.305\linewidth]{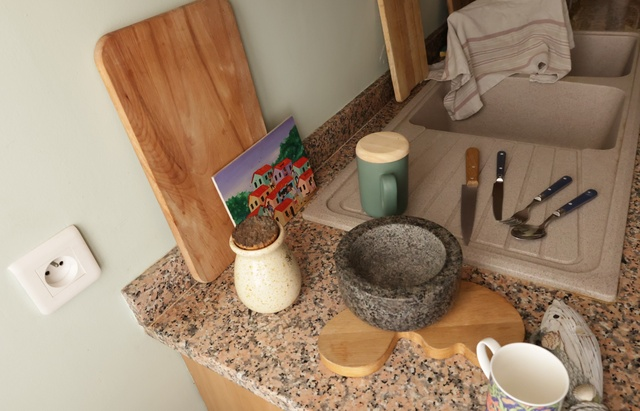} & 
\includegraphics[width=.302\linewidth]{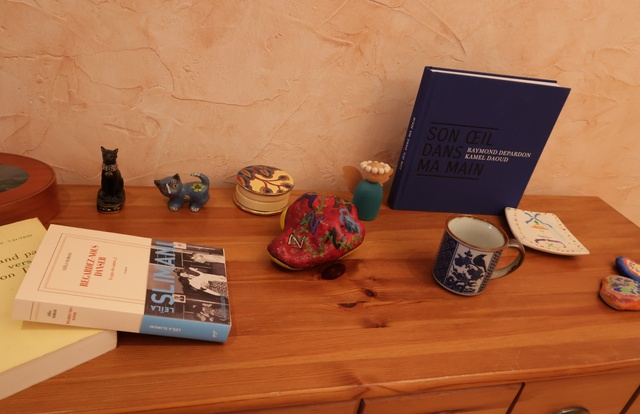} \\

{\rotatebox[origin=lB]{90}{\hspace{0.5in}\small{Re-render}}} & 
\includegraphics[width=.352\linewidth]{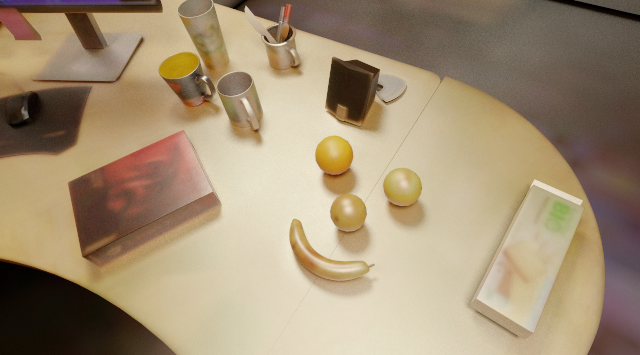} & 
\includegraphics[width=.305\linewidth]{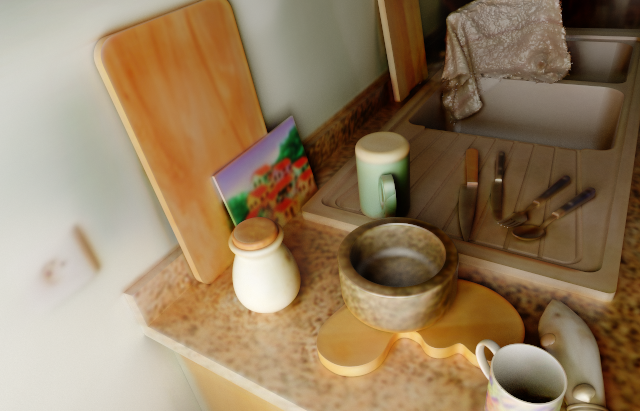} & 
\includegraphics[width=.302\linewidth]{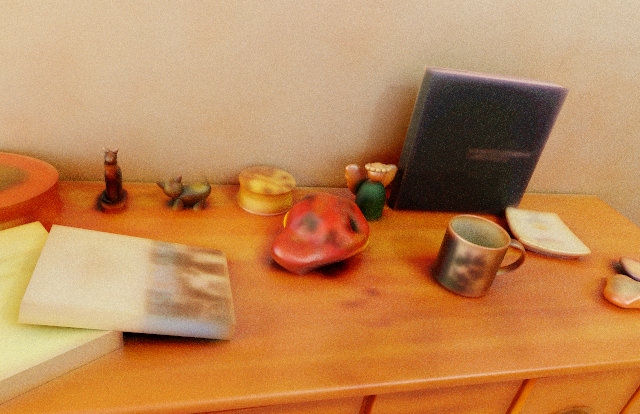} \\

{\rotatebox[origin=lB]{90}{\hspace{0.5in}\small{Diffuse}}} & 
\includegraphics[width=.352\linewidth]{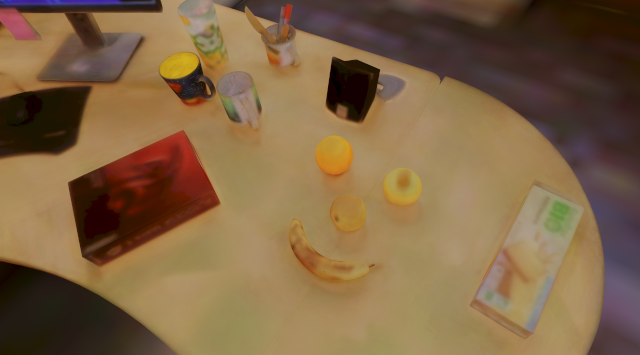} & 
\includegraphics[width=.305\linewidth]{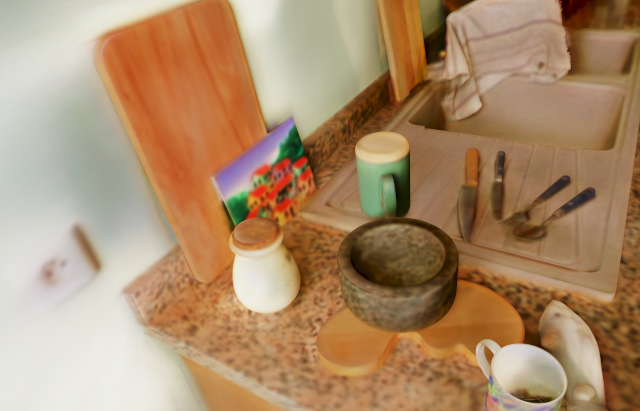} & 
\includegraphics[width=.302\linewidth]{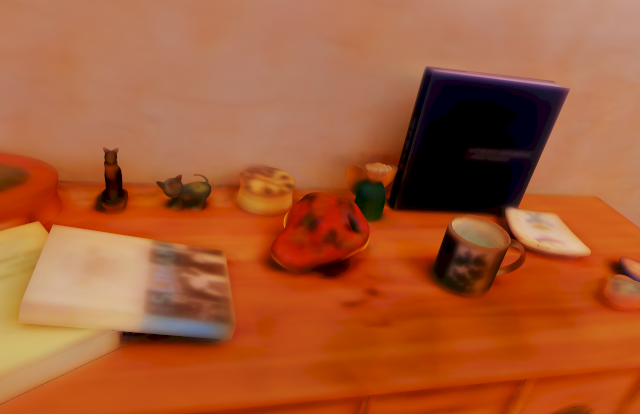} \\

{\rotatebox[origin=lB]{90}{\hspace{0.3in}\small{Roughness}}} & 
\includegraphics[width=.352\linewidth]{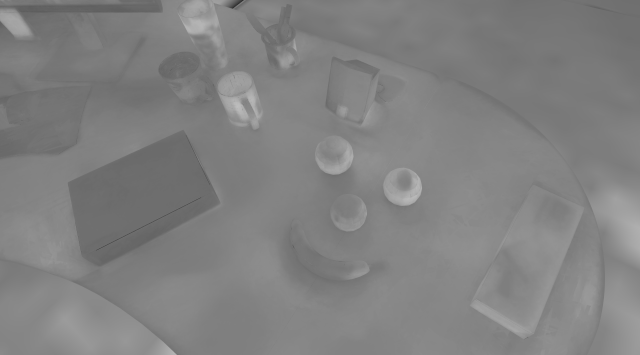} & 
\includegraphics[width=.305\linewidth]{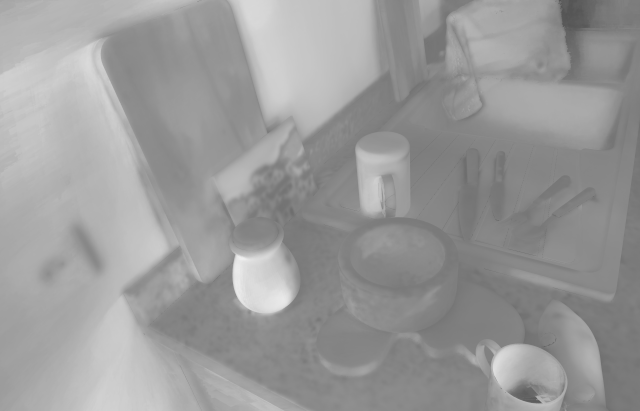} & 
\includegraphics[width=.302\linewidth]{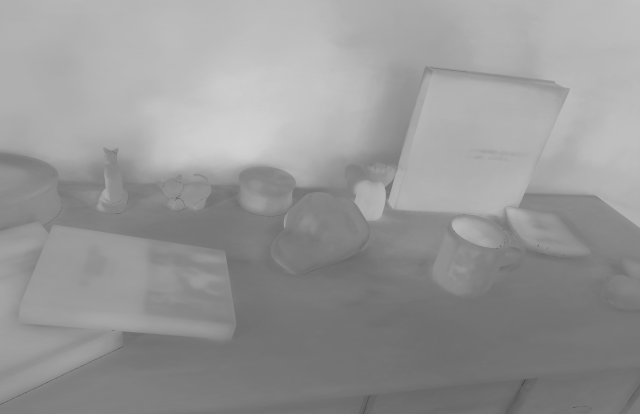} \\

{\rotatebox[origin=lB]{90}{\hspace{0.5in}\small{Specular}}} & 
\includegraphics[width=.352\linewidth]{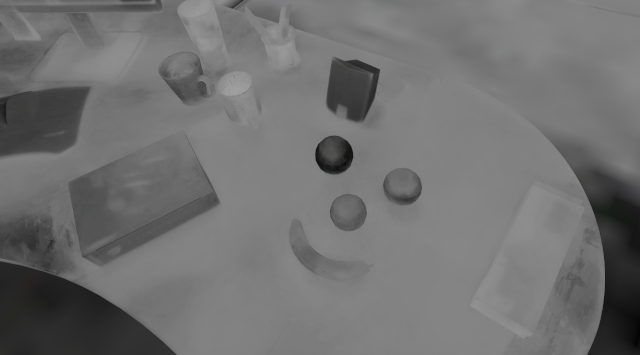} &
\includegraphics[width=.305\linewidth]{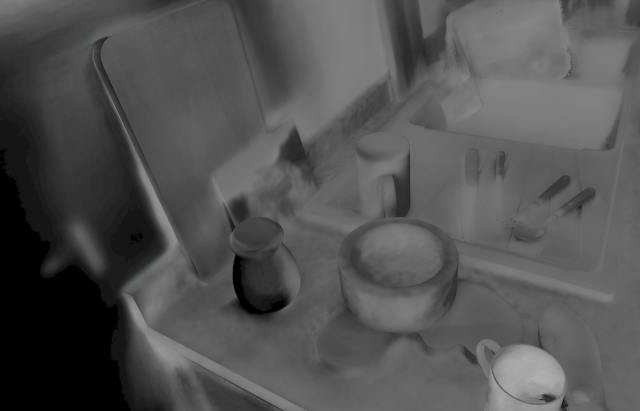} &
\includegraphics[width=.302\linewidth]{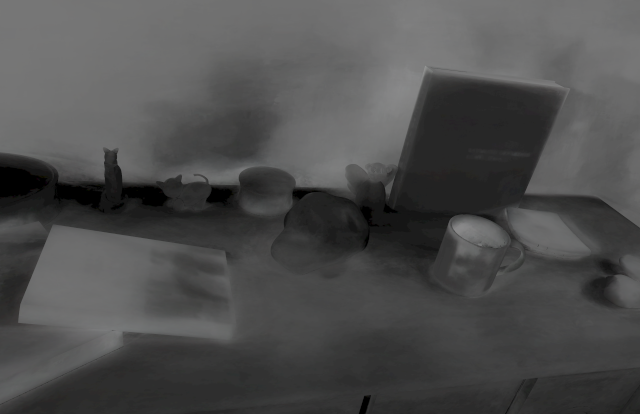} \\
%\hline

\end{tabular}
\caption{
\label{fig:results-real}
Results on real scenes. For each scene, we show the re-rendering in closest matched capture conditions and the maps obtained for the given input view using our method.
}
\end{figure*}

\begin{figure*}[h!]
\setlength{\tabcolsep}{1pt}
\begin{tabular}{ccccc}
%\hline
 & Capture Configuration & Modified Lighting & +~Object Insertion &  +~Novel View  \\
%\hline
{\rotatebox[origin=lB]{90}{\hspace{0.5in}\small{\textbf{Real Office}}}} & 
\begin{tikzpicture}[x=4cm, y=7cm, spy using outlines={width=2cm, height=2cm, rectangle,magnification=2}]
\node[anchor=south] (FigA) at (0,0) {
	\includegraphics[width=.23\linewidth]{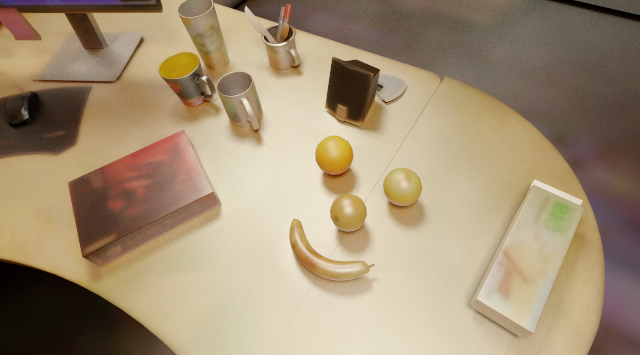} 
};
\spy [smallwindow] on ($(FigA)+(-0.18,- 0.02)$) in node[largewindow,anchor=north west] at ($(FigA.south west)+(0.035,0.00)$);
\spy [smallwindow4] on ($(FigA)+(0.02, 0.04)$) in node[largewindow4,anchor=north east] at ($(FigA.south east)+(-0.035,0.00)$);
\end{tikzpicture}& 
\begin{tikzpicture}[x=4cm, y=7cm, spy using outlines={width=2cm, height=2cm, rectangle,magnification=2}]
\node[anchor=south] (FigA) at (0,0) {
	\includegraphics[width=.23\linewidth]{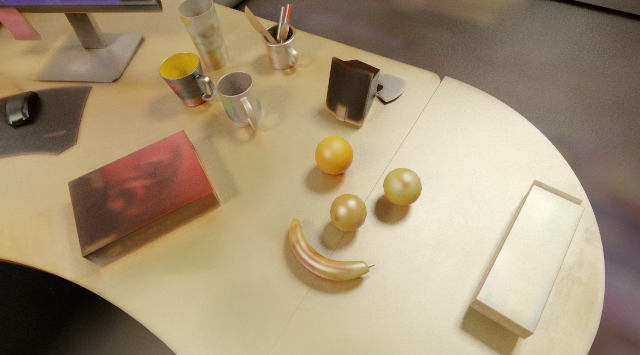}
};
\spy [smallwindow] on ($(FigA)+(-0.18, -0.02)$) in node[largewindow,anchor=north west] at ($(FigA.south west)+(0.035,0.00)$);
\spy [smallwindow4] on ($(FigA)+(0.02, 0.04)$) in node[largewindow4,anchor=north east] at ($(FigA.south east)+(-0.035,0.00)$);
\end{tikzpicture} & 
\begin{tikzpicture}[x=4cm, y=7cm, spy using outlines={width=2cm, height=2cm, rectangle,magnification=2}]
\node[anchor=south] (FigA) at (0,0) {
	\includegraphics[width=.23\linewidth]{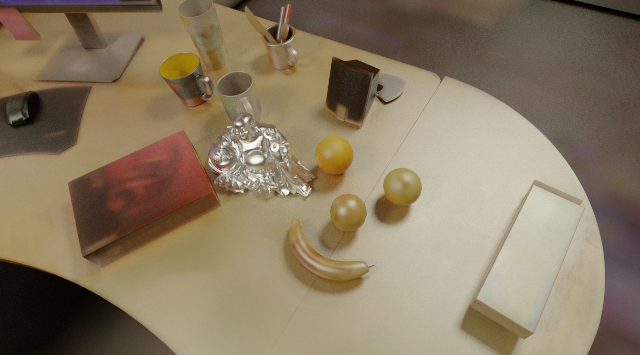}
};
\spy [smallwindow] on ($(FigA)+(-0.18, -0.02)$) in node[largewindow,anchor=north west] at ($(FigA.south west)+(0.035,0.00)$);
\spy [smallwindow4] on ($(FigA)+(0.02, 0.04)$) in node[largewindow4,anchor=north east] at ($(FigA.south east)+(-0.035,0.00)$);
\end{tikzpicture} & 
\begin{tikzpicture}[x=4cm, y=7cm, spy using outlines={width=2cm, height=2cm, rectangle,magnification=3}]
\node[anchor=south] (FigA) at (0,0) {
	\includegraphics[width=.23\linewidth]{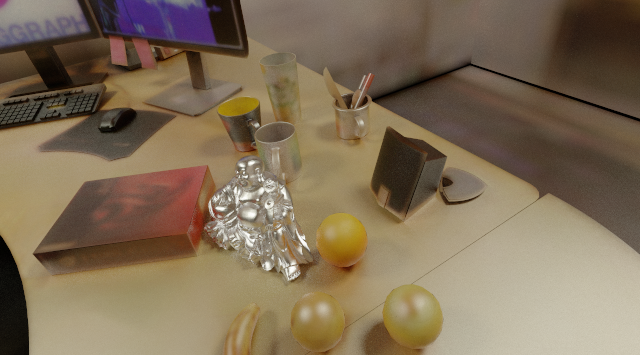} 
};
\spy [smallwindow2] on ($(FigA)+(-0.2, -0.03)$) in node[largewindow2,anchor=north west] at ($(FigA.south west)+(0.035,0.00)$);
\spy [smallwindow3] on ($(FigA)+(0.0, -0.07)$) in node[largewindow3,anchor=north east] at ($(FigA.south east)+(-0.035,0.00)$);
\end{tikzpicture}  \\

%\hline
{\rotatebox[origin=lB]{90}{\hspace{0.5in}\small{\textbf{Real Kitchen}}}} & 
\begin{tikzpicture}[x=4cm, y=7cm, spy using outlines={width=2cm, height=2cm, rectangle,magnification=2}]
\node[anchor=south] (FigA) at (0,0) {
	\includegraphics[width=.23\linewidth]{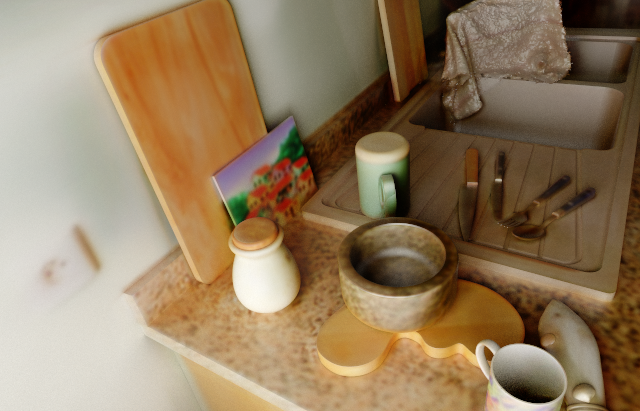} 
};
\spy [smallwindow] on ($(FigA)+(-0.12, 0.05)$) in node[largewindow,anchor=north west] at ($(FigA.south west)+(0.035,0.00)$);
\spy [smallwindow4] on ($(FigA)+(0.25, 0.0)$) in node[largewindow4,anchor=north east] at ($(FigA.south east)+(-0.035,0.00)$);
\end{tikzpicture}& 
\begin{tikzpicture}[x=4cm, y=7cm, spy using outlines={width=2cm, height=2cm, rectangle,magnification=2}]
\node[anchor=south] (FigA) at (0,0) {
	\includegraphics[width=.23\linewidth]{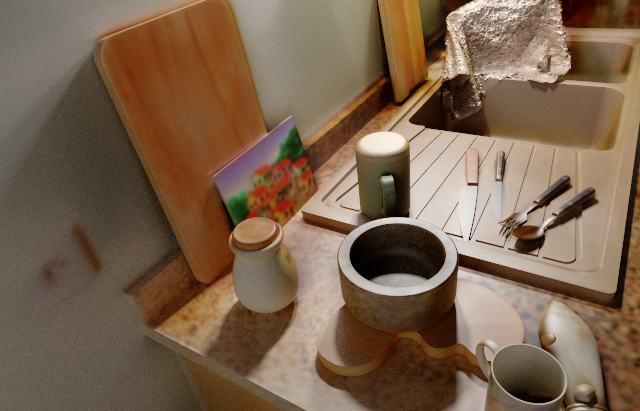}
};
\spy [smallwindow] on ($(FigA)+(-0.12, 0.05)$) in node[largewindow,anchor=north west] at ($(FigA.south west)+(0.035,0.00)$);
\spy [smallwindow4] on ($(FigA)+(0.25, 0.0)$) in node[largewindow4,anchor=north east] at ($(FigA.south east)+(-0.035,0.00)$);
\end{tikzpicture} & 
\begin{tikzpicture}[x=4cm, y=7cm, spy using outlines={width=2cm, height=2cm, rectangle,magnification=2}]
\node[anchor=south] (FigA) at (0,0) {
	\includegraphics[width=.23\linewidth]{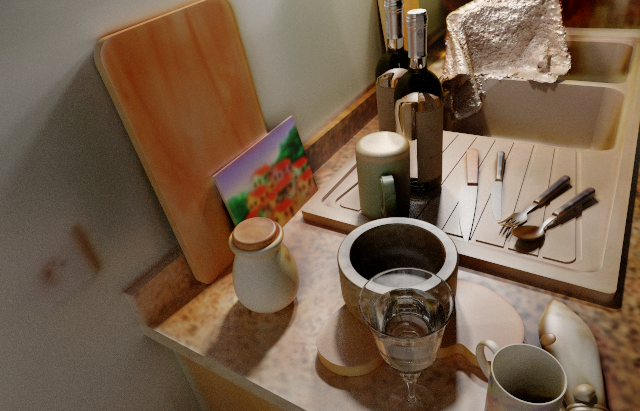}
};
\spy [smallwindow] on ($(FigA)+(-0.12, 0.05)$) in node[largewindow,anchor=north west] at ($(FigA.south west)+(0.035,0.00)$);
\spy [smallwindow4] on ($(FigA)+(0.25, 0.0)$) in node[largewindow4,anchor=north east] at ($(FigA.south east)+(-0.035,0.00)$);
\end{tikzpicture} & 
\begin{tikzpicture}[x=4cm, y=7cm, spy using outlines={width=2cm, height=2cm, rectangle,magnification=3}]
\node[anchor=south] (FigA) at (0,0) {
	\includegraphics[width=.23\linewidth]{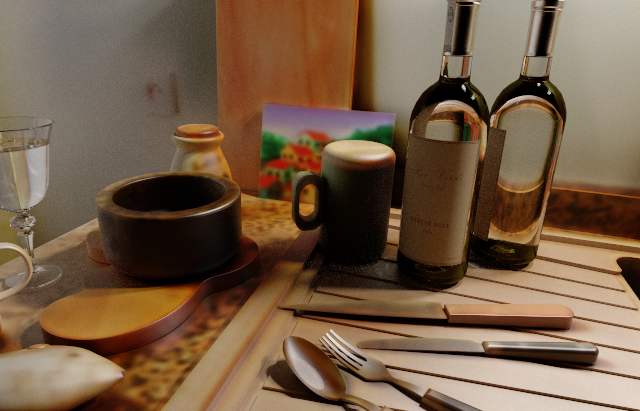} 
};
\spy [smallwindow2] on ($(FigA)+(-0.45, 0.03)$) in node[largewindow2,anchor=north west] at ($(FigA.south west)+(0.035,0.00)$);
\spy [smallwindow3] on ($(FigA)+(0.24, 0.08)$) in node[largewindow3,anchor=north east] at ($(FigA.south east)+(-0.035,0.00)$);
\end{tikzpicture}  \\

%\hline
{\rotatebox[origin=lB]{90}{\hspace{0.5in}\small{\textbf{Real Hallway}}}} & 
\begin{tikzpicture}[x=4cm, y=7cm, spy using outlines={width=2cm, height=2cm, rectangle,magnification=2}]
\node[anchor=south] (FigA) at (0,0) {
	\includegraphics[width=.23\linewidth]{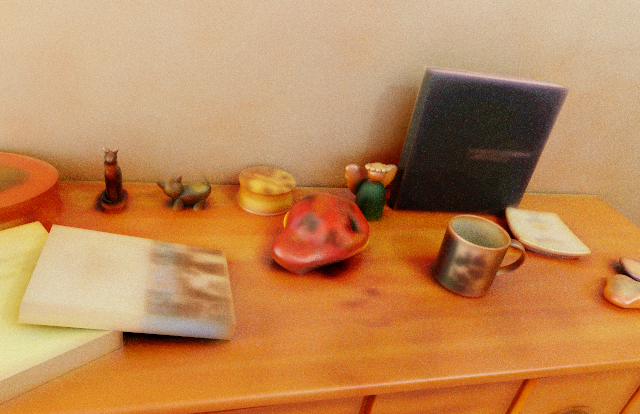} 
};
\spy [smallwindow] on ($(FigA)+(-0.06,- 0.05)$) in node[largewindow,anchor=north west] at ($(FigA.south west)+(0.035,0.00)$);
\spy [smallwindow4] on ($(FigA)+(0.12, 0.08)$) in node[largewindow4,anchor=north east] at ($(FigA.south east)+(-0.035,0.00)$);
\end{tikzpicture}& 
\begin{tikzpicture}[x=4cm, y=7cm, spy using outlines={width=2cm, height=2cm, rectangle,magnification=2}]
\node[anchor=south] (FigA) at (0,0) {
	\includegraphics[width=.23\linewidth]{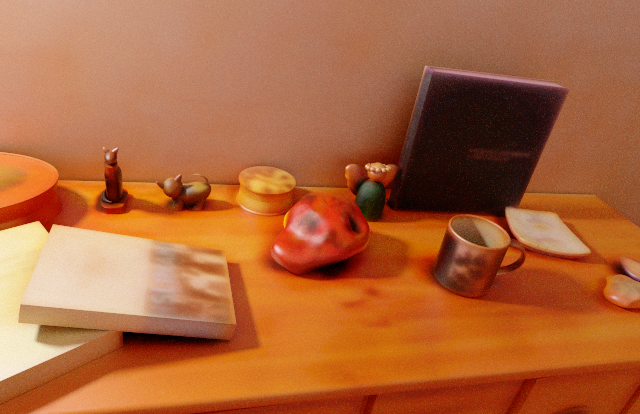}
};
\spy [smallwindow] on ($(FigA)+(-0.06,- 0.05)$) in node[largewindow,anchor=north west] at ($(FigA.south west)+(0.035,0.00)$);
\spy [smallwindow4] on ($(FigA)+(0.12, 0.08)$) in node[largewindow4,anchor=north east] at ($(FigA.south east)+(-0.035,0.00)$);
\end{tikzpicture} & 
\begin{tikzpicture}[x=4cm, y=7cm, spy using outlines={width=2cm, height=2cm, rectangle,magnification=2}]
\node[anchor=south] (FigA) at (0,0) {
	\includegraphics[width=.23\linewidth]{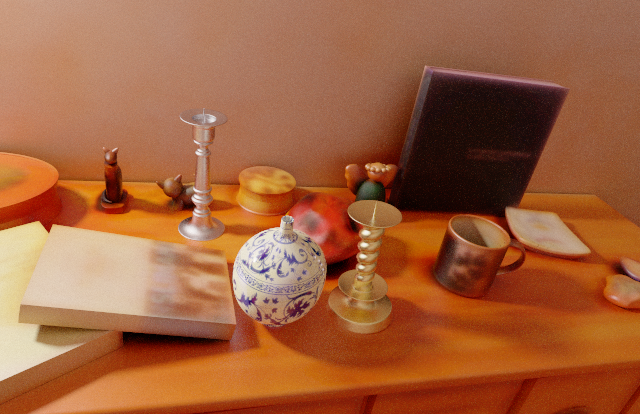}

};
\spy [smallwindow] on ($(FigA)+(-0.06,- 0.05)$) in node[largewindow,anchor=north west] at ($(FigA.south west)+(0.035,0.00)$);
\spy [smallwindow4] on ($(FigA)+(0.12, 0.08)$) in node[largewindow4,anchor=north east] at ($(FigA.south east)+(-0.035,0.00)$);
\end{tikzpicture} & 
\begin{tikzpicture}[x=4cm, y=7cm, spy using outlines={width=2cm, height=2cm, rectangle,magnification=3}]
\node[anchor=south] (FigA) at (0,0) {
	\includegraphics[width=.23\linewidth]{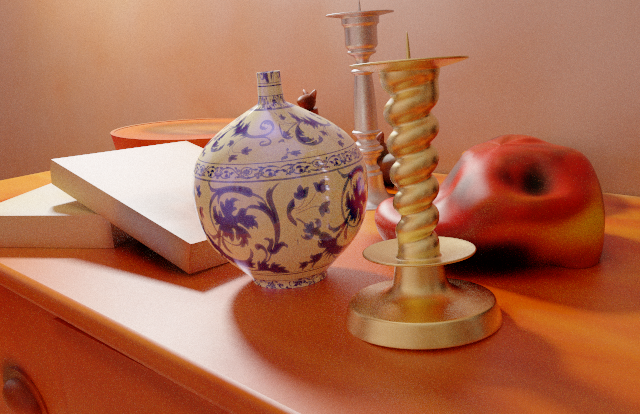} 

};
\spy [smallwindow2] on ($(FigA)+(-0.0, 0.03)$) in node[largewindow2,anchor=north west] at ($(FigA.south west)+(0.035,0.00)$);
\spy [smallwindow3] on ($(FigA)+(0.08, -0.02)$) in node[largewindow3,anchor=north east] at ($(FigA.south east)+(-0.035,0.00)$);
\end{tikzpicture}  \\

%\hline
\end{tabular}
\caption{
\label{fig:results2}
Scene editing with our method on real scenes from input and novel viewpoints.
From left to right, we show the scene re-rendered with capture configurations, the same view re-rendered with modified lighting, with a virtual object inserted along with modified lights, and the modified scene from a novel viewpoint which is never captured. 
For example, for the real office scene we have added the statuette; notice  how shadows underneath the red book and orange fruit have been (re)moved correctly on modifying lights (first row, blue and red inset) and the color bleeding of the red box and orange on the statuette (first row, green and yellow inset) due to \emph{correct interaction between real and virtual objects} which cannot be achieved by relighting based object-insertion methods. 
Our method enables insertion of objects with highly complex materials such as in the real kitchen scene, we have inserted a goblet of water and two bottles of wine. Since our method produces digital assets that can be directly used for rendering, our renderings exhibit complex lighting phenomena such as caustics (second row, red inset) and internal reflection (second row, green and yellow inset) which is enabled by our method while removing shadows (second row, blue inset; third row red inset) and correctly predicting the specular material e.g., on the knives and the red shoe-stone in real hallway scene as illustrated by the rendered highlights (second row, red inset; third row, blue inset).
Please see the supplementary video for results with moving paths and lights to better appreciate these effects. 
}
\end{figure*}

\section{Implementation Details}
\label{sec:implementation}
We have implemented our method in python using the PyTorch~\cite{pytorch} framework for deep learning and C++/OpenGL shaders for all steps that require reprojection. 

At inference, we run the CNN over images in $640 \times 384$ resolution. If the aspect ratio does not match, we zero-pad to fit to the nearest multiple resolution. 
We expect input images in linear color space, and we apply a log transformation followed by a normalization to $[-1, 1 ]$ to flatten the dynamic range before processing by the CNN. 
For real images, we 
assume gamma correction of $2.2$ to convert to linear space. 

Our system for dataset generation includes a plug-in for 3DS Max that exports materials into \textsc{Mitsuba}-compatible format using our BRDF model. 
To handle complex material graphs, we evaluate them in 3DS Max and save texture layers subsequently used by \textsc{Mitsuba}. 
%\GD{ADD DETAILS HERE on choices of parameters etc}

%\AB{Provide details about training (learning rate, epoch, training time)}
We first train the diffuse track and then the specular track separately over 15 epochs each, using ground truth maps for the missing components when evaluating the rendering loss.
We then fine-tune the two tracks jointly for 15 epochs. Overall training takes 18 hours on a 4 RTX8000 GPU cluster node.
The confidence networks for roughness and specular maps are trained for $100$ iterations.
We use the \textsc{Adam} optimizer~\cite{kingmaba15} with a fixed learning rate of $2e-5$ for all training.

We will release all source code for our method, including the dataset generation system, and all training images used for our results.

%We first train the diffuse albedo track to get a good baseline for underlying surface color/texture and then train the roughness/specular maps track\AB{while holding the albedo track fixed? Does it matter?}, followed by a joint fine-tuning step (Fig.~\ref{fig:overview}(b). For individual track training, for example when training only the diffuse track, we use predicted albedo along with ground truth specular and roughness maps to render the predicted image while we use all three ground truth maps for ground truth rendering.

%\begin{itemize}
%		\item For inference we provide images in 640x384 resolution. If the aspect ration does not match, we pad by zeros to fit to the nearest multiple resolution. 
%		\item All inputs are provided in linear space. For real images, we apply gamma correction of 2.2.
%		\item We obtain pre-view output maps in linear space. 
%		\item For a dataset of 310 images, it takes about 5-10 minutes to infer all maps through our prediction network. 
%\end{itemize}

%\begin{table}[!t]
%%\setlength{\tabcolsep}{2pt}
%\begin{tabular}{l||ccc}
%%\hline
%Scenes & \#images & Resolution & Lighting condition \\
%\hline \hline
%Veach Ajar & $160$ & $640 \times 384$ & $3$~area \\
%Dining Room & $105$ & $640 \times 384$ & $2$~area + 1 envmap \\
%Real Office & $245$ & $640 \times 355$ & $2$~area \\
%Real Kitchen & $296$ & $640 \times 411$ & $1$~area \\
%Real Hallway & $226$ & $640 \times 414$ & $2$~area + 1 lamp \\
%\hline \hline
%\end{tabular}
%\caption{
%\label{tab:scenes}
%\NEW{Details of scenes used for our experiments.}}
%\end{table}

\begin{table*}[!t]
\centering
\begin{tabular}{l||cccccc}
%\hline
Scenes & \#images & Resolution & Pre-processing & Prediction & Texturing ($\times 3$) & Total  \\
\hline \hline
Veach Ajar &  $160$ & $640 \times 384$ & $7.56$ & $3.93$ & $3.12$ & $20.85$ \\
Dining Room & $105$ & $640 \times 384$ & $4.56$ & $3.26$ & $1.32$ & $9.14$ \\
Real Office & $245$ & $640 \times 355$ & $11.48$ & $5.84$ & $4.77$ & $31.63$ \\
Real Kitchen & $296$ & $640 \times 411$ & $32.05$ & $25.25$ & $7.91$ & $81.03$ \\
Real Hallway & $226$ & $640 \times 414$ & $16.10$ & $10.26$ & $7.03$ & $47.45$ \\
\hline \hline
Mean & & & $14.35$ & $9.71$ & $4.83$ & $\textbf{38.02}$ \\
\hline \hline
\end{tabular}
\caption{
\label{tab:timings}
Timing breakdown for each step of our method on the scenes used for our experiments. All times are reported in \textbf{minutes}.}
\end{table*}

\section{Results and Evaluation}
\label{sec:experiments}

We show results and evaluations on two synthetic scenes (Veach Ajar, and  Dining Room), and three real captured scene (Real Office, Real Kitchen, and Real Hallway). 
The capture details and scene lighting conditions are provided in Table~\ref{tab:timings}.
We also provide comparisons on two additional synthetic scenes (Living Room, and Kitchen).
We provide 
%supplemental webpage with all results and comparisons as well as 
a supplemental video showing view dependent effects over paths and image sequences.
We strongly encourage the reader to view the videos to appreciate how our automatically created material maps are directly usable for physically-based rendering and scene editing.

For synthetic scenes, we render a set of views of the scene, and then use these as if they were photographs to run our entire pipeline; we use the original geometry in this case. 
Note that as a result the ground truth materials are encoded as a single material map texture atlas to be comparable with the results of our method. %\SP{Unclear to me what the last sentence conveys.}
For real scenes, we take a set of photos of the scene, paying attention to capture highlights in several views, then run structure-from-motion and multi-view stereo to obtain an initial 3D reconstruction. 
We hired a professional artist to turn these reconstructions into a retopologized mesh (shown in Fig.~\ref{fig:overview}(a) and~\ref{fig:geometry}).
We use Blender automatic UV-unwrapper to unwrap the meshes into texture atlas.
%% GD: prefer not to mention, will tell at rebuttal if asked , which took around \AB{how many?} hours per scene. 
The resolution of the texture atlas is 16Kx16K pixels for the scenes we considered.
%, followed by a retopology step performed by an artist. 

Synthetic data allows quantitative comparisons on both the material maps and the renderings; for real scenes we can only show qualitative results due to the lack of ground truth maps.
%compute quantitative results only on re-renderings of the input views. Note however that the retoplogized geometry does not match the images \emph{exactly} and thus there are reprojection and misalignment errors that occur.
%\GD{Maybe held out from inference, although it doesnt really matter} 
%\SP{We can't really perform quantitative on real scenes even with held-out views since the underlying geometry model is not same as the real one. Maybe it is worth mentioning this distinction in our work from IBR.}

\subsection{Results}
\label{sec:results}

%		\item Augmentations - Real Scenes re-rendered with:
%		\begin{enumerate}
%			\item Different illuminations: demonstrate shadow/highlight removal
%			\item Different materials: demonstrate material editing
%			\item Object insertion: demonstrate shadows, highlights, inter-reflections, caustics 
%		\end{enumerate}

In Fig.~\ref{fig:results1} and \ref{fig:results-real}, 
%we show an input view from each scene (a), 
%visualization of the maps on the (retopologized) geometry (b)-(d) and the re-rendered image (e) in the first row, 
%the corresponding maps from the per-view inference, (b)-(d) in the second row and the
%ground truth maps and re-rendering of our final method in the third row for each synthetic scene.
%we show the scene geometry textured with maps estimated from a single camera versus the final maps merged in texture space.
we show the scene geometry textured with final \emph{approximate} material maps obtained by our method.
For synthetic scenes we also show the ground truth maps.
Additionally, for both synthetic and real scenes,  we show the re-rendered image, i.e., we generate the material map texture atlases using our method and then provide them to a path tracer along with the geometry to render the scene with full global illumination effects.
To achieve a result as close as possible to the input image, 
we place the lights manually to match input conditions as much as possible. 
Despite these approximations, our re-renderings are plausible renditions of the input images, illustrating the efficacy of the approximate material maps we obtain.

%Although, due to the inherent difference between actual scene lighting and the approximate geometry from the input, the global illumination effects are not 100\% accurate but still we are able to reproduce images that are very similar to the input images showing the efficacy of the approximate material maps we obtain.} 

%as well as the re-rendering image using our rendering pipeline, i.e., using the \emph{per-view} ground truth maps to create a texture atlas and then rendering using the same method as with inference. The difference from the input is mainly due to global illumination effects from parts of the scene not seen in the input views that do not have materials assigned (i.e., are black). 

%For synthetic scenes, we see in Fig.~\ref{fig:results1},\ref{fig:results-real} that our method predicts reasonable values for the maps compared to the ground truth and results in a re-rendering that is quite close to the ground truth (far right top), and the input. 

Our method manages to capture the overall material properties of the objects even in real scenes, e.g., in the Real Office scene the desktop and the red box are shiny while the yellow box on the right and the orange are more diffuse. 

In Fig.~\ref{fig:results2} we show results with modified lighting conditions and object insertion from different viewpoints on real scenes.
This figure shows that we achieve our goal of creating plausible material assets for photorealistic scene editing.
For each scene, we show a view in input lighting condition and the same view with modified lighting condition. We are able to remove and move shadows and highlights on most surfaces. 
We further augment the scene by inserting complex objects, such as a metallic statuette in the Real Office scene and a transparent water goblet and wine bottles in the Real Kitchen scene.
Our material maps, together with the retopologized geometry are complete digital assets, and thus allow renderings with full GI interactions between real and virtual objects, 
%provide information about the underlying materials of surrounding objects, these virtual objects interact realistically with the real objects and are able to exhibit complex light phenomena 
such as color bleeding, refraction, caustics and internal reflections when rendered using a path tracer.
We emphasize that such effects are not possible to reproduce using relighting based object insertion methods such as~\cite{karsch2011, karsch2014, gardner2017}.
%Finally we also show the edited scene (with modified lighting and inserted objects) from a novel viewpoint which is never captured before and no material maps are predicted by our network for this particular viewpoint. %% GD said in the beginning

\subsection{Evaluation}
\label{sec:evaluation}

The only other methods designed to handle our input at scene-scale are differentiable rendering approaches~\cite{merlin2021, haefner2021}; unfortunately neither code nor data (in the form of input images we can use for SfM/MVS) is available, precluding direct comparison.
In any case, our method can be seen as complementary and could be used as an initialization for these methods, potentially accelerating their process.
We report the timings for different steps of our method and compare with timings reported by these previous works to support our claims.
Additionally, we present best-effort evaluation using two baselines. 
We also present a set of ablation studies to analyse the effect of our various design choices.

\subsubsection{Speed}
We show the timing breakdown of each step for our method (pre-processing (\textsc{Pre-processing}), single-view prediction (\textsc{Prediction}), and texture atlas generation (\textsc{Texturing}) on a system with an Intel Xeon Gold 5218 2.30GHz Processor and Quadro RTX 5000 GPU, in Tab.~\ref{tab:timings}.
%we need $\sim8 min$ to perform inference for a typical dataset of around 300 images, while reprojection into texture space requires $\sim15 min$. Pre-processing to compute the statistics or re-projected colors into each input view requires approximately $\sim15-30 min$. 
We observe that it is possible to create renderable assets from a multi-view dataset with approximately $30~minutes$ to an hour of computation depending on number of images and their resolution. 
Previous works dealing with scene-scale material estimation~\cite{merlin2021, haefner2021} report around $10-12$ hours for a complete scene optimization. 
Thus, our method is able to produce renderable material maps in \emph{a fraction of time} as compared to previous works.
Note that the timings reported do not include the time taken for reconstruction and re-topology of the mesh.
% since we do not consider it a part of the pipeline.}
%\SP{I have moved timings from implementation details to here.}

%\begin{table}[!t]
%\setlength{\tabcolsep}{2pt}
%\begin{tabular}{l||cc|cc}
%\hline
% & \multicolumn{2}{c|}{\textbf{Veach Ajar}} & \multicolumn{2}{c}{\textbf{Dining Room}}\\
%Method & PSNR $\uparrow$ & DSSIM $\downarrow$ & PSNR $\uparrow$ & DSSIM $\downarrow$ \\
%\hline \hline
%\textsc{ \small{ \textsc{LiEtAl} } } &\small{12.974275} & \small{0.370910} & \small{21.335057} &\small{0.359676}\\
%\textsc{ \small{ \textsc{Texture} } } &\small{18.701869} & \small{0.303566} & \small{15.123681}& \small{0.618277} \\
%\textsc{ \small{ Ours } } & \textbf{\small{21.533800}} & \textbf{\small{0.266334}} & \textbf{\small{21.773710}} & \textbf{\small{0.338615}} \\
%\hline \hline
%\end{tabular}
%\caption{
%\label{tab:comparisons}
% Quantitative comparison. Our method performs best for both Veach Ajar and Dining Room scenes which supports our qualitative observations~(see also Fig.~\protect{\ref{fig:comparisons1}}).
%}
%\end{table}

\begin{table}[!t]
\setlength{\tabcolsep}{2pt}
\begin{tabular}{l||ccc||ccc}
%\hline\hline
 & \multicolumn{3}{c||}{PSNR $\uparrow$} & \multicolumn{3}{c}{DSSIM $\downarrow$}\\
Method & \textsc{LiEtAl} & \textsc{Texture} & \textsc{Ours} & \textsc{LiEtAl} & \textsc{Texture} & \textsc{Ours} \\
\hline \hline
\textbf{Veach Ajar} &\small{12.97} &  \small{18.70} & \textbf{\small{21.53}} & \small{0.37} &\small{0.30} & \textbf{\small{0.27}}\\
\textbf{Dining Room} & \small{21.33}  & \small{15.12}& \textbf{\small{21.77}}&\small{0.36} & \small{0.62}  & \textbf{\small{0.34}} \\
\textbf{Living Room} & \small{13.67}  & \small{19.97}& \textbf{\small{25.61}}&\small{0.44} & \small{0.44}  & \textbf{\small{0.16}} \\
%\textbf{Bathroom} & \small{14.12}  & \small{14.83}& \textbf{\small{20.27}}&\small{0.41} & \small{0.43}  & \textbf{\small{0.32}} \\
\textbf{Kitchen} & \small{9.06}  & \small{16.54}& \textbf{\small{21.28}}&\small{0.52} & \small{0.42}  & \textbf{\small{0.30}} \\
\hline\hline
\textbf{Mean} & \small{14.26}  & \small{17.58}& \textbf{\small{22.55}}&\small{0.42} & \small{0.44}  & \textbf{\small{0.27}} \\
\hline \hline
\end{tabular}
\caption{
\label{tab:comparisons}
%\TODO{RECOMPUTE NUMBERS}
 Quantitative comparison on 4 synthetic scenes. Our method performs best across the scenes which supports our qualitative observations~(see also Fig.~\protect{\ref{fig:comparisons1} and \ref{fig:comparisons3}}).
}
\end{table}

\begin{figure*}[!h]
\setlength{\tabcolsep}{1pt}
\begin{tabular}{ccccc}
%\hline
& \multicolumn{2}{c}{\textbf{Synthetic Veach Ajar}} & \multicolumn{2}{c}{\textbf{Synthetic Dining Room}}\\
%\hline
& Input View & Novel View & Input View & Novel View \\
%\hline
{\rotatebox[origin=lB]{90}{\small{Original Light GT}}} & 
\includegraphics[width=.24\linewidth]{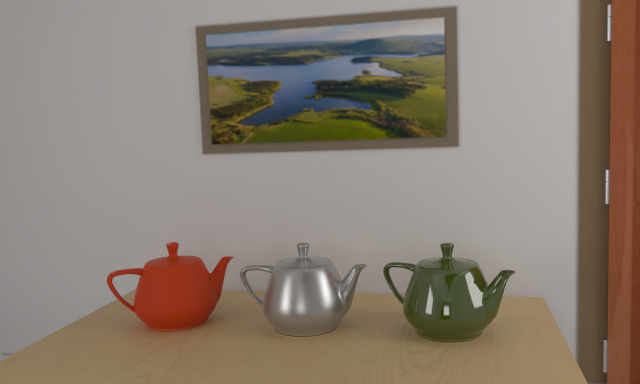} & 
\includegraphics[width=.24\linewidth]{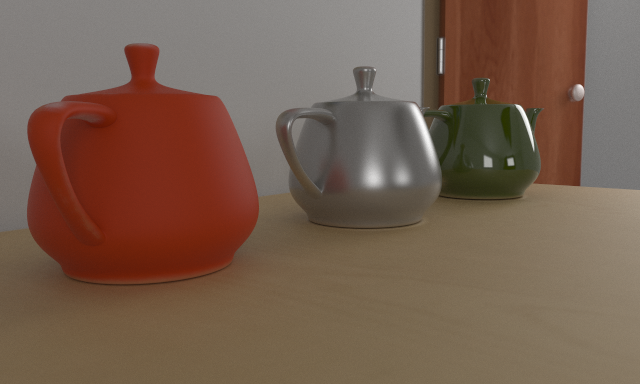} &
\includegraphics[width=.24\linewidth]{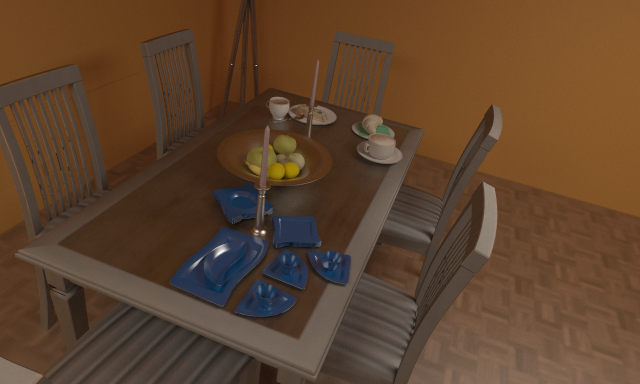} & 
\includegraphics[width=.24\linewidth]{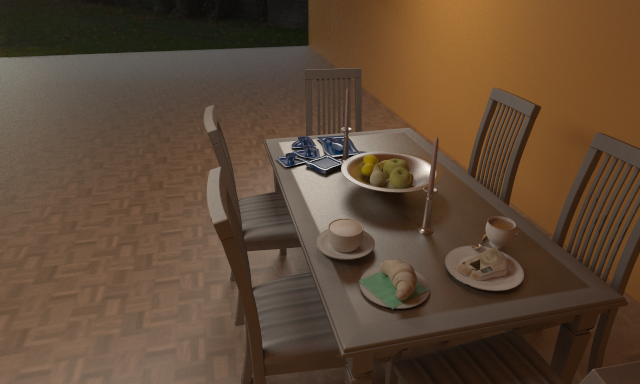} \\

{\rotatebox[origin=lB]{90}{\small{Modified Light GT}}} &
\includegraphics[width=.24\linewidth]{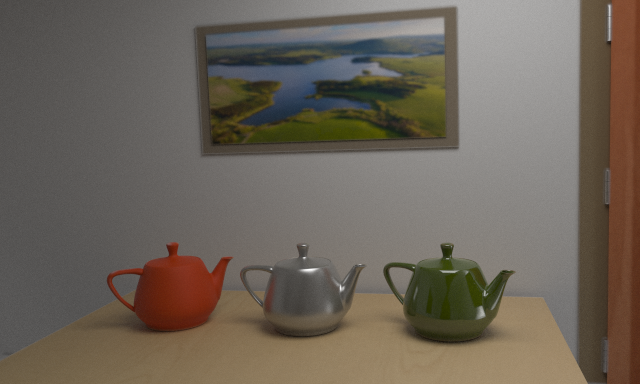} & 
\includegraphics[width=.24\linewidth]{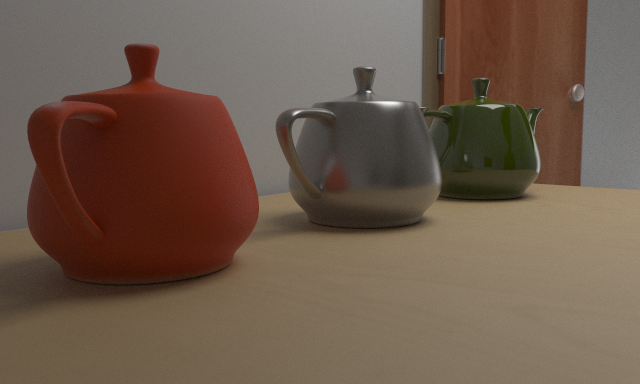} &
\includegraphics[width=.24\linewidth]{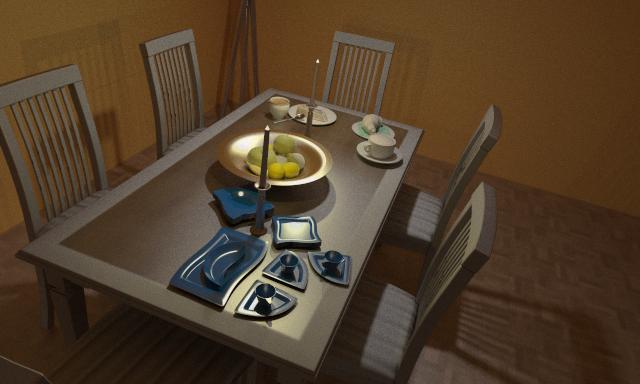} & 
\includegraphics[width=.24\linewidth]{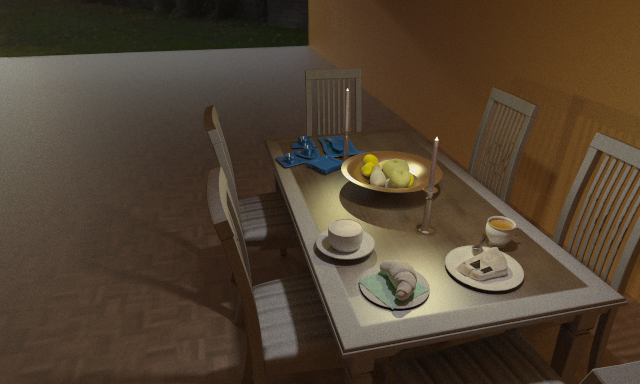} \\

{\rotatebox[origin=lB]{90}{\hspace{0.35in}\small{Ours}}} &
\includegraphics[width=.24\linewidth]{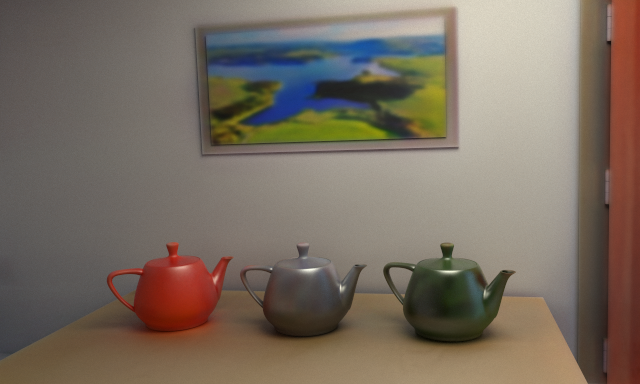} & 
\includegraphics[width=.24\linewidth]{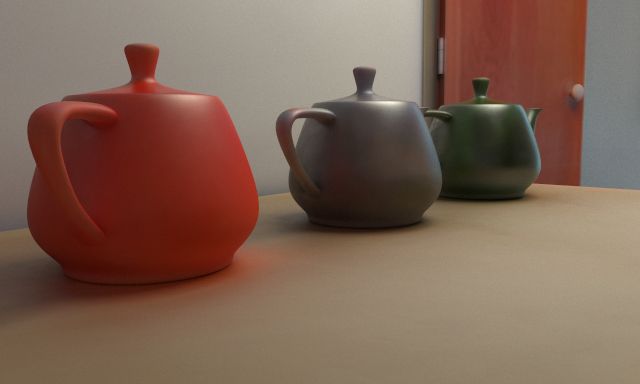} &
\includegraphics[width=.24\linewidth]{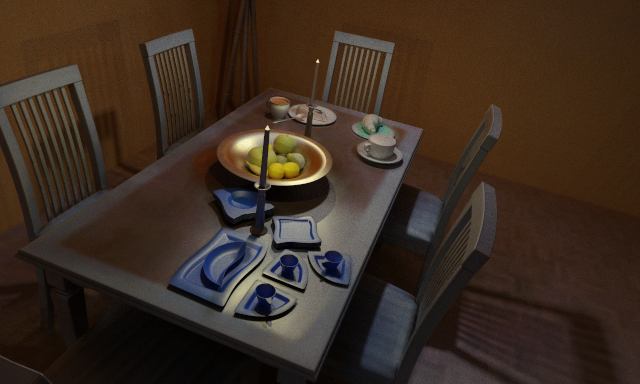} & 
\includegraphics[width=.24\linewidth]{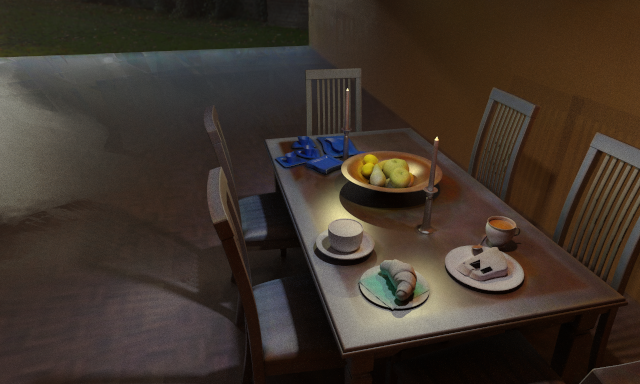} \\

{\rotatebox[origin=lB]{90}{\hspace{0.25in}\small{\textsc{LiEtAl}}}} &
\includegraphics[width=.24\linewidth]{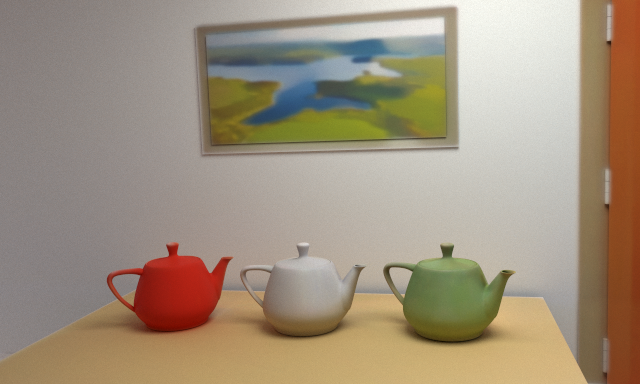} & 
\includegraphics[width=.24\linewidth]{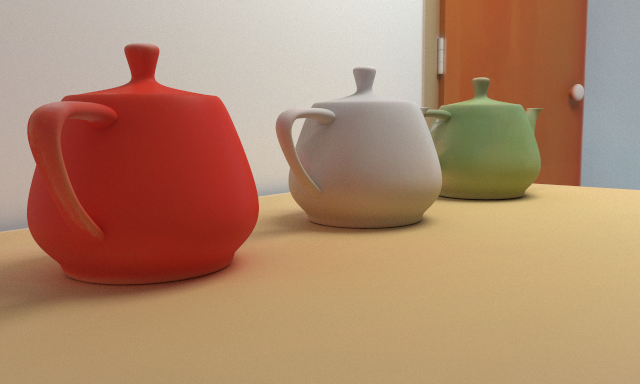} &
\includegraphics[width=.24\linewidth]{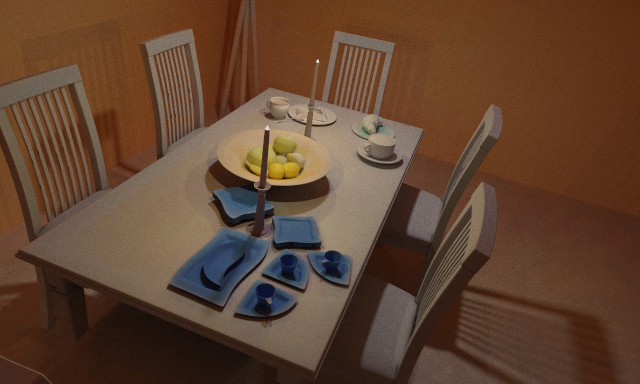} & 
\includegraphics[width=.24\linewidth]{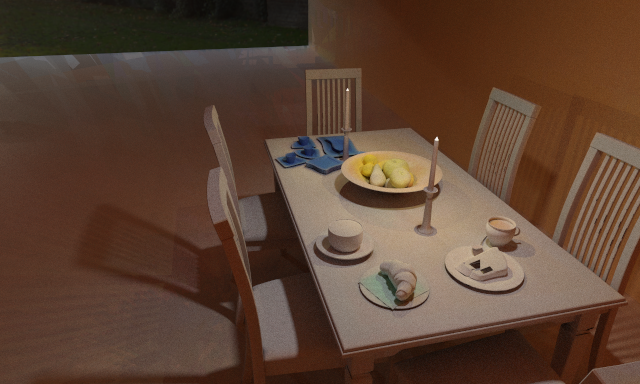} \\

{\rotatebox[origin=lB]{90}{\hspace{0.25in}\small{\textsc{Texture}}}} &
\includegraphics[width=.24\linewidth]{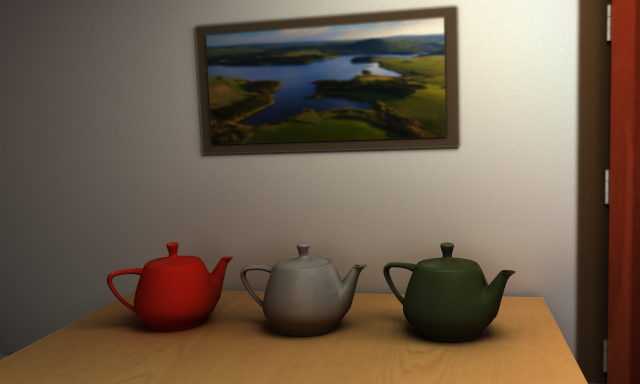} &
\includegraphics[width=.24\linewidth]{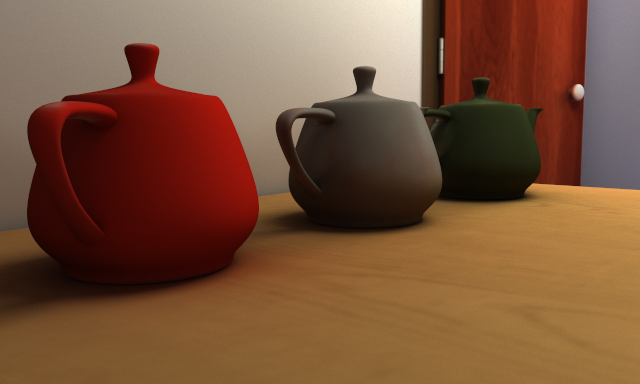} &
\includegraphics[width=.24\linewidth]{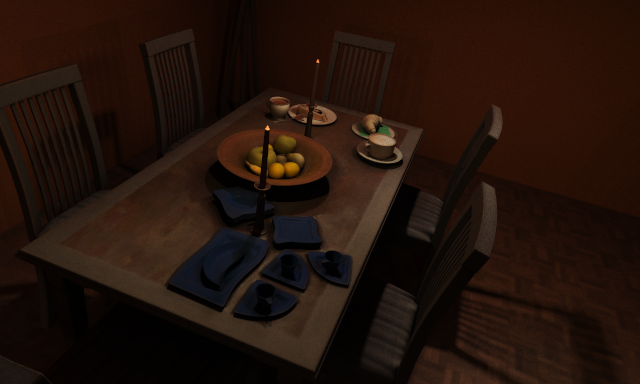} &
\includegraphics[width=.24\linewidth]{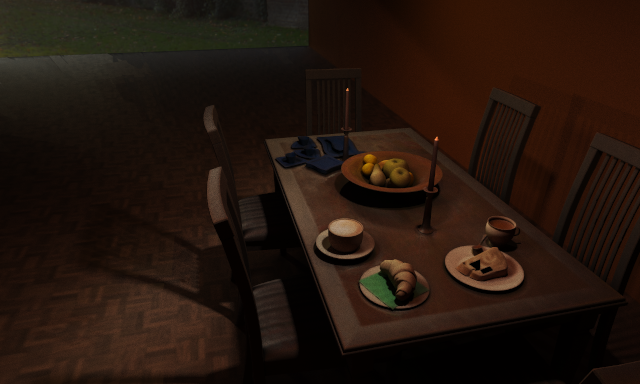} \\ 

%\hline
\end{tabular}
\caption{
\label{fig:comparisons1}
Comparisons with baselines on synthetic scenes;
For each scene, we show an input view and a novel view; in the first row we show the viewpoint with original input lighting condition (\textsc{Original Light GT}). 
From second row onwards we modify the lighting in the scene and show the same view re-rendered with modified lighting condition using ground truth maps (\textsc{Modified Light GT}), our maps (\textsc{Ours}), maps produced by \textsc{LiEtAl} and the baseline static \textsc{Texture} generated using the input images. 
Column 1 and 3 corresponds to an input view while column 2 and 4 corresponds to a novel view for which no maps were predicted. 
Notice how \textsc{LiEtAl} fails to reconstruct highlight properly and the \textsc{Texture} is composed of pre-baked highlights and shadows from original lighting condition.
}
\end{figure*}

\begin{figure*}[!h]
\setlength{\tabcolsep}{1pt}
\begin{tabular}{ccccccc}
%\hline
&  \multicolumn{2}{c}{\textbf{Synthetic Living Room}}  &  \multicolumn{2}{c}{\textbf{Synthetic Kitchen}}\\
%\hline
& Input View & Novel View &  Input View & Novel View\\
%\hline
{\rotatebox[origin=lB]{90}{\small{Original Light GT}}} & 

\includegraphics[width=.24\linewidth]{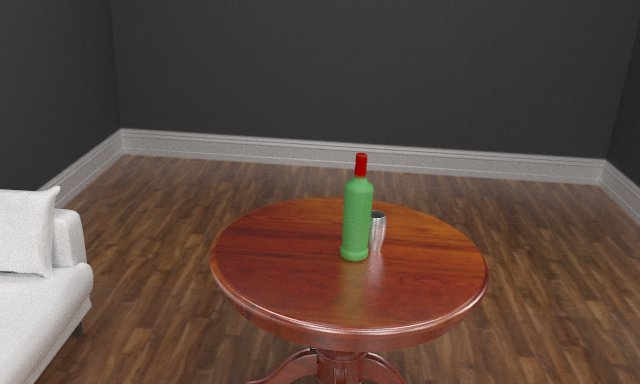} & 
\includegraphics[width=.24\linewidth]{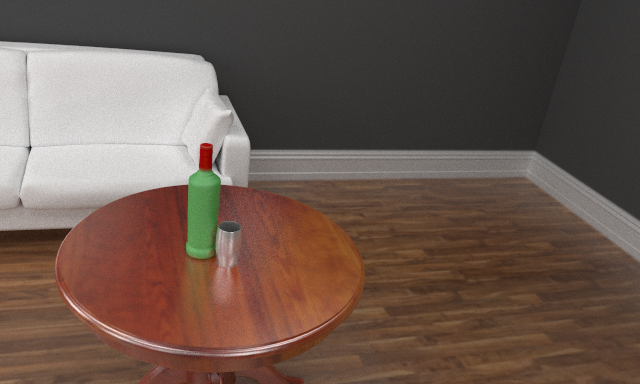} &
\includegraphics[width=.24\linewidth]{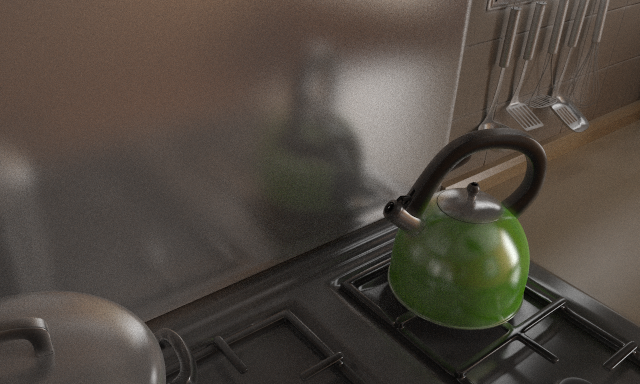} &
\includegraphics[width=.24\linewidth]{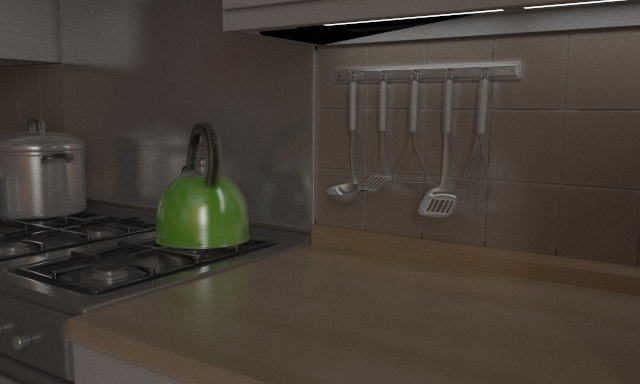} \\

{\rotatebox[origin=lB]{90}{\small{Modified Light GT}}} &

\includegraphics[width=.24\linewidth]{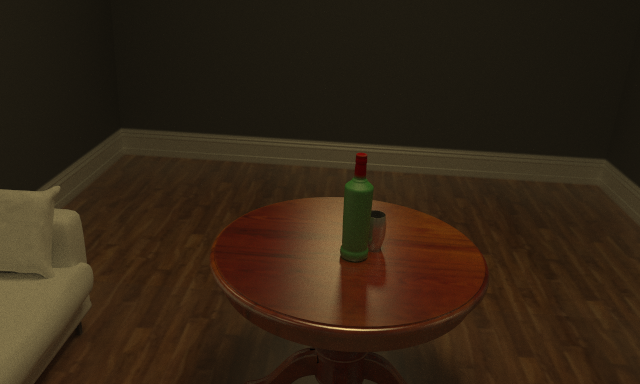} & 
\includegraphics[width=.24\linewidth]{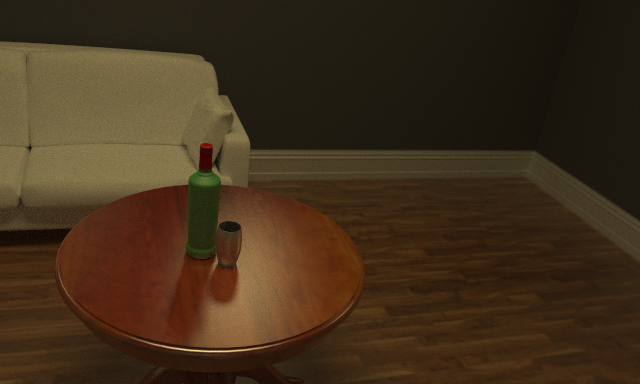} &
\includegraphics[width=.24\linewidth]{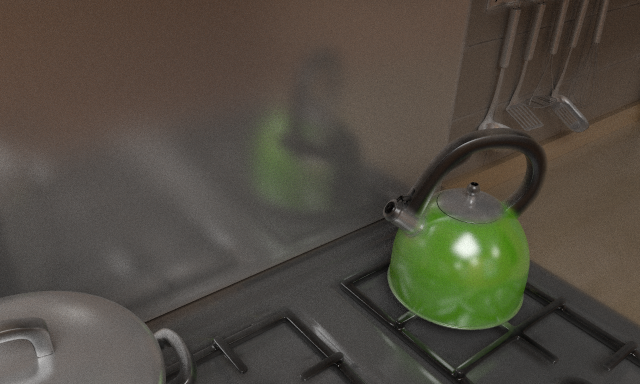} &
\includegraphics[width=.24\linewidth]{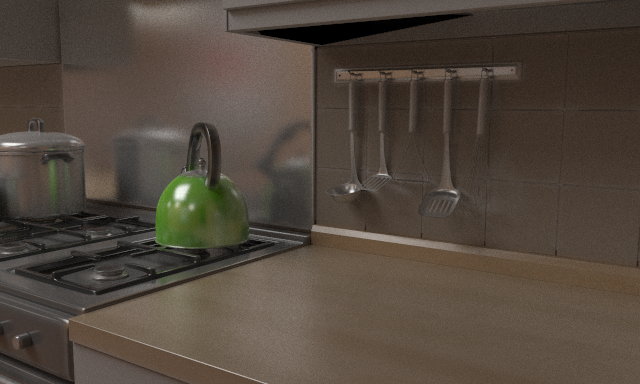} \\

{\rotatebox[origin=lB]{90}{\hspace{0.35in}\small{Ours}}} &
\includegraphics[width=.24\linewidth]{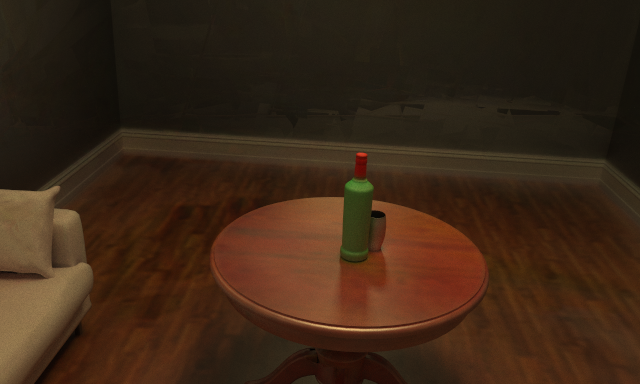} & 
\includegraphics[width=.24\linewidth]{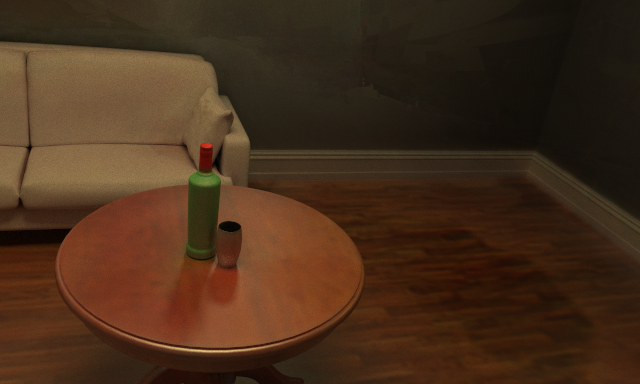} &
\includegraphics[width=.24\linewidth]{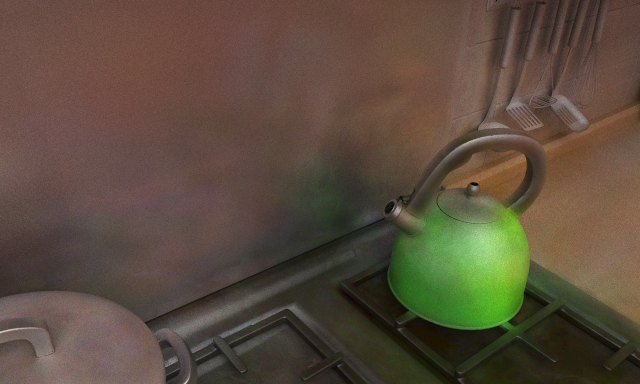} &
\includegraphics[width=.24\linewidth]{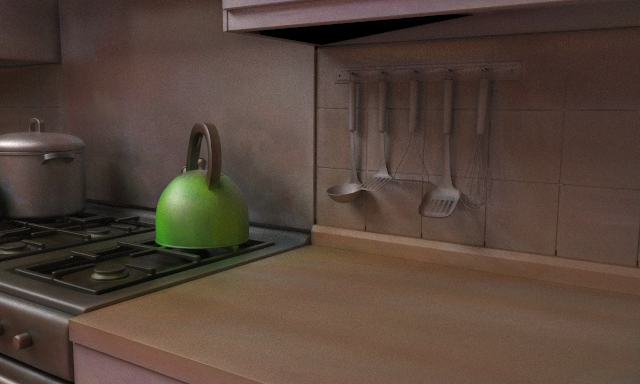} \\

{\rotatebox[origin=lB]{90}{\hspace{0.25in}\small{\textsc{LiEtAl}}}} &
\includegraphics[width=.24\linewidth]{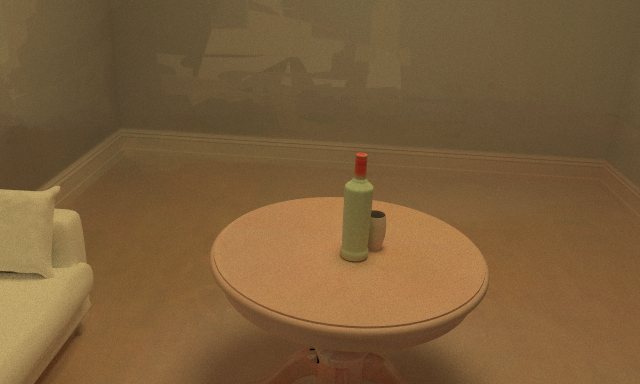} & 
\includegraphics[width=.24\linewidth]{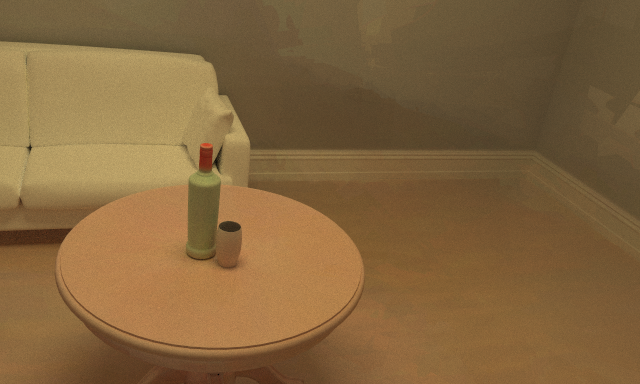} &
\includegraphics[width=.24\linewidth]{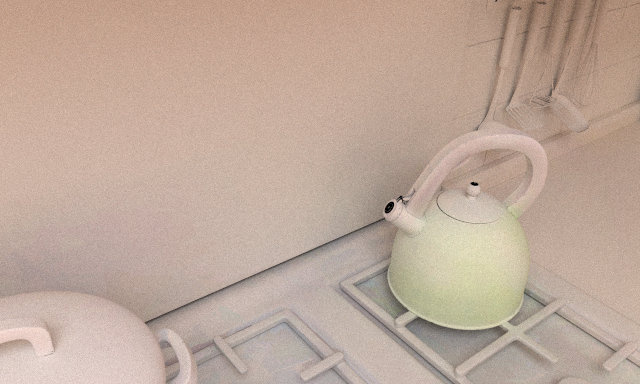} &
\includegraphics[width=.24\linewidth]{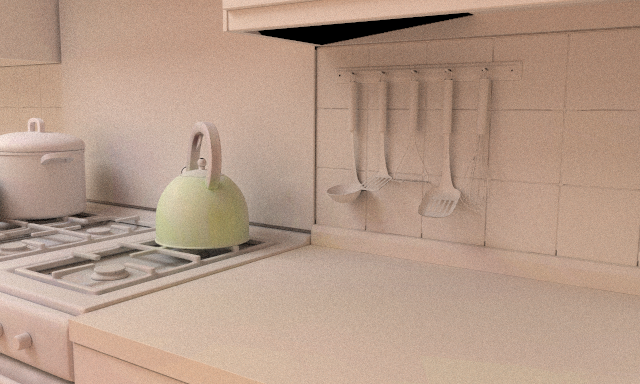} \\

{\rotatebox[origin=lB]{90}{\hspace{0.25in}\small{\textsc{Texture}}}} &
\includegraphics[width=.24\linewidth]{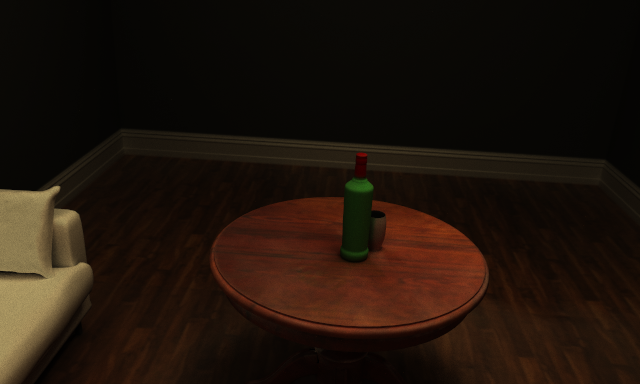} &
\includegraphics[width=.24\linewidth]{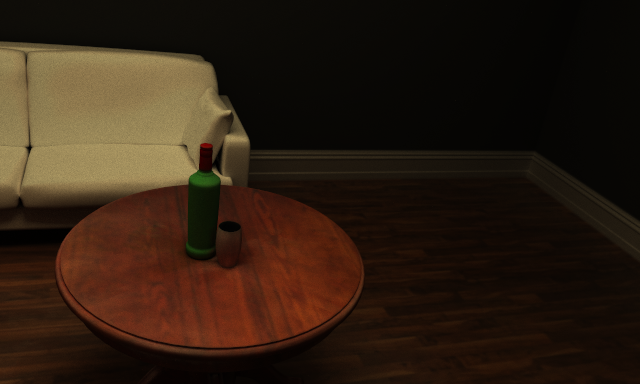} &
\includegraphics[width=.24\linewidth]{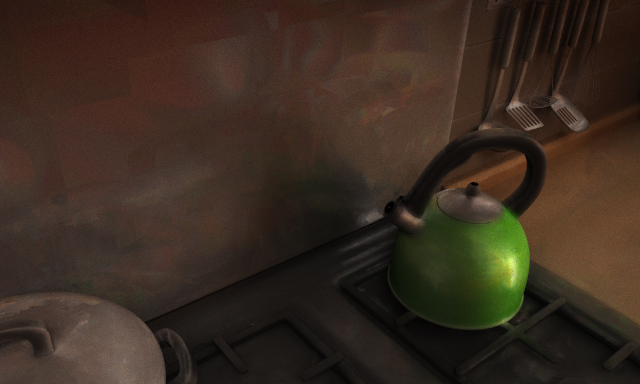} &
\includegraphics[width=.24\linewidth]{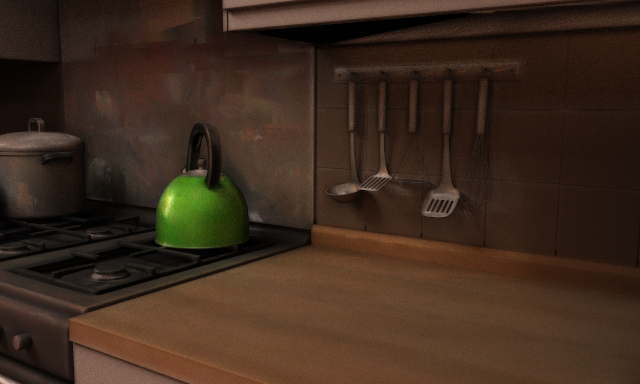} \\ 

%\hline
\end{tabular}
\caption{
\label{fig:comparisons3}
More comparisons with baselines on synthetic scenes;
The layout is the same as for Fig.~\ref{fig:comparisons1}.
Similarly as in Fig.~\ref{fig:comparisons1} we see that our approach shows much better variations of appearance, i.e., shiny materials, compared to the \textsc{LiEtAl} that struggles with shiny material sand texture details and \textsc{Texture} that contains pre-baked highlights and shadows from input images owing to its diffuse and static nature.
%For each scene, we show an input view and a novel view; in the first row we show the viewpoint with original input lighting condition (\textsc{Original Light GT}). 
%From second row onwards we modify the lighting in the scene and show the same view re-rendered with modified lighting condition using ground truth maps (\textsc{Modified Light GT}), our maps (\textsc{Ours}), maps produced by \textsc{LiEtAl} and the baseline static \textsc{Texture} generated using the input images. 
%Column 1 and 3 corresponds to an input view while column 2 and 4 corresponds to a novel view for which no maps were predicted. 
%Notice how \textsc{LiEtAl} fails to reconstruct highlight properly and the \textsc{Texture} is composed of pre-baked highlights and shadows from original lighting condition.
}
\end{figure*}

\subsubsection{Comparisons}

We compare to two baselines: 
%the first uses the retopologized geometry and performs median-filtered texture mapping (\textsc{Texture}), 
The first mimics current practice in digital content creation and uses the retopologized geometry to project the input images into texture space, using median filtering to create an RGB atlas (\textsc{Texture}), 
while the second is based on the single image method of Li et al.~\cite{li2020invrender} (\textsc{LiEtAl}). Specifically, we run the method of Li and colleagues to generate maps for each input view, then we run the same pipeline as for our method to create a material texture atlas for the scene from the multiple predicted maps. We implement their BRDF model in \textsc{Mitsuba} to generate the re-renderings.

We perform quantitative and qualitative comparisons for our method compared to the two baselines. Since the first baseline does not estimate maps, and \textsc{LiEtAl} infers maps for a different BRDF model, we only compare re-renderings, i.e., we re-render the scene for a set of views. 
We perform quantitative comparisons by computing PSNR and DSSIM error~\cite{loza2006structural}.
We show the numerical results in Tab.~\ref{tab:comparisons}, and a visual comparison in Fig.~\ref{fig:comparisons1} and \ref{fig:comparisons3}. 
For the visual comparison, we show the input view on which the maps are predicted. In line with our aim of generating plausible material assets for scene editing, we generate ground truth re-renderings of a modified scene by rendering the scene with modified lighting using ground truth SVBRDF maps. To show how our method compares against the two baseline for this task, we re-render the scene with modified lighting but with the material maps obtained by the three methods.
We see that our approach shows much better variations of appearance, i.e., shiny materials, compared to the \textsc{LiEtAl} that struggles to identify shiny materials and \textsc{Texture} where everything is diffuse by construction.
Our method works well for novel views which we show by re-rendering a novel view for each scene. This view was never seen by the network and no maps were predicted for this viewpoint.
%For the visual comparison, we show the input image, and a re-rendering of the same viewpoint with our method, \textsc{LiEtAl} and \textsc{Texture}.
The quantitative results are computed on $10$ views of each scene selected from a rendered path; we also show a video comparison on the rendered path along with the ground truth in the supplemental video. 
We see in Tab.~\ref{tab:comparisons} that our method is numerically best for both synthetic scenes on both PSNR and DSSIM metrics. 
In Veach Ajar scene our method performs significantly better than the baselines as we can see how our method is able to approximately recover the gradation in specularity between the teapots while both the baselines fails to do so.
In Dining Room scene, numerically our results are better yet close to \textsc{LiEtAl}, although it's worth noticing that visually our method is able to do much better. For example, we are able to recover the spatial variation in roughness of the table top which \textsc{LiEtAl} fails to do and ends up over-smoothing the diffuse albedo. This shows the advantage of using multi-view information which helps the network infer the spatially-varying nature of roughness locally as well as globally for each surface point.

\begin{figure*}[!h]
\setlength{\tabcolsep}{1pt}
\begin{tabular}{cccccc}
%\hline
 & Frame 1 & Frame 2 & Frame 3 & Frame 4 & Frame 5 \\
%\hline
{\rotatebox[origin=lB]{90}{\hspace{0.2in}\small{Ours}}} &
\includegraphics[width=.19\linewidth]{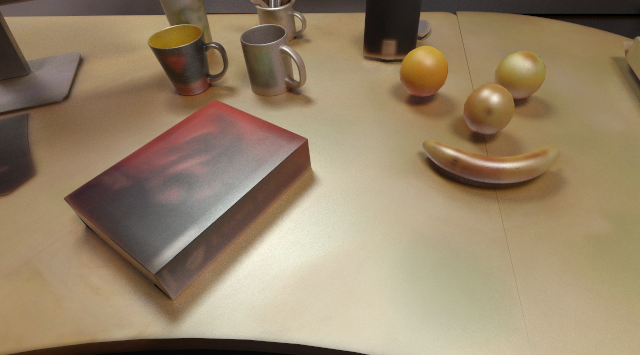} & 
\includegraphics[width=.19\linewidth]{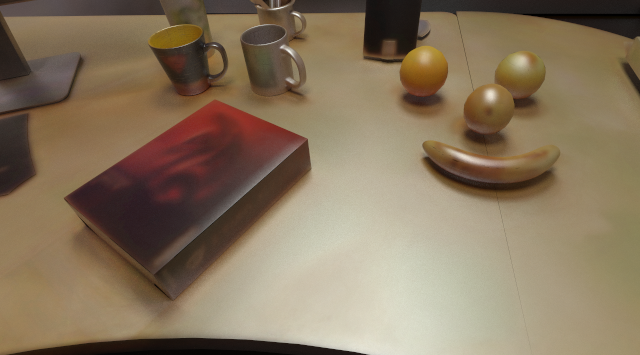} & 
\includegraphics[width=.19\linewidth]{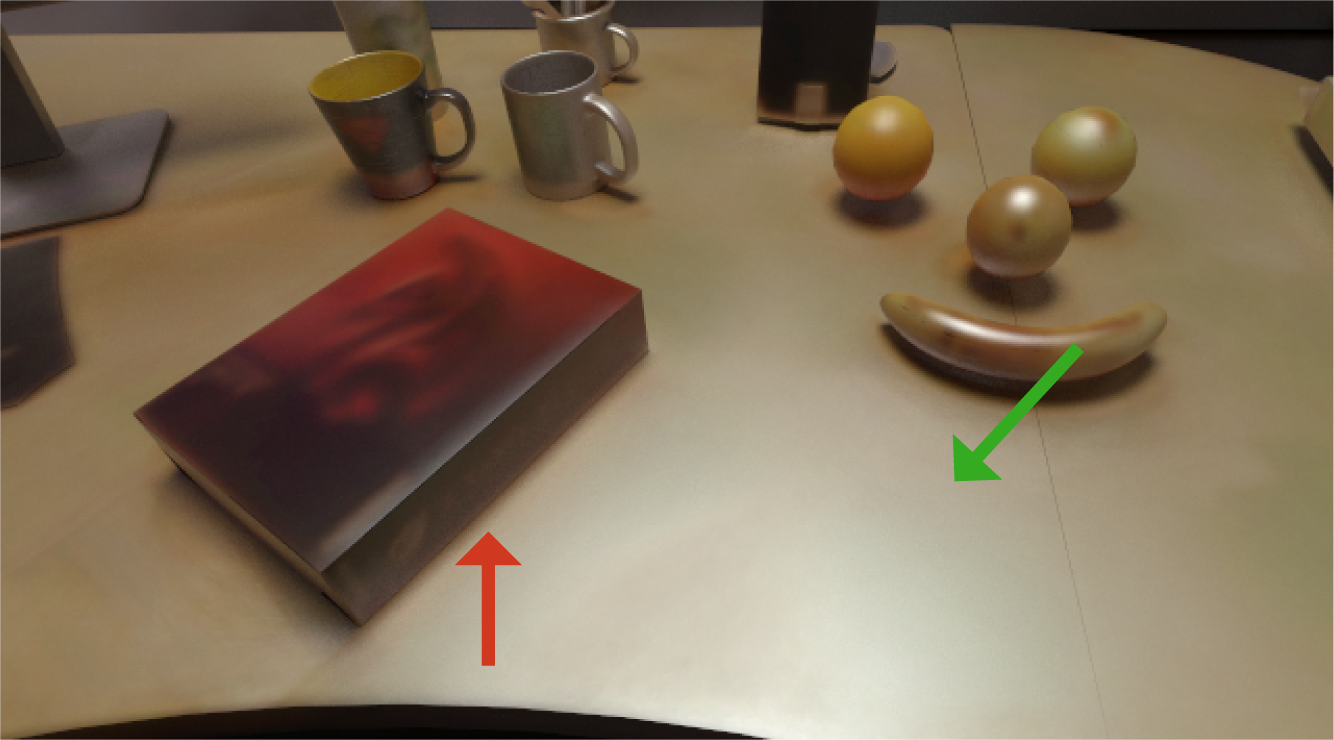} & 
\includegraphics[width=.19\linewidth]{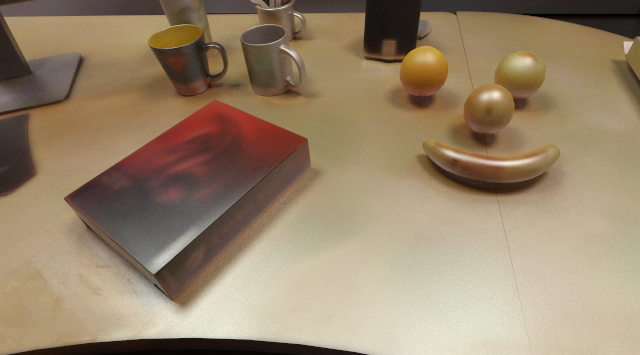} & 
\includegraphics[width=.19\linewidth]{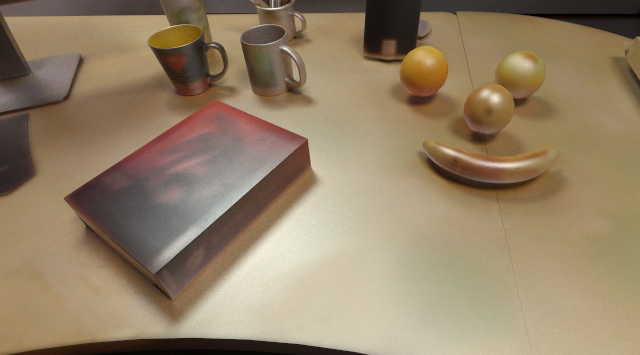} \\ 
{\rotatebox[origin=lB]{90}{\hspace{0.2in}\small{\textsc{LiEtAl}}}} &
\includegraphics[width=.19\linewidth]{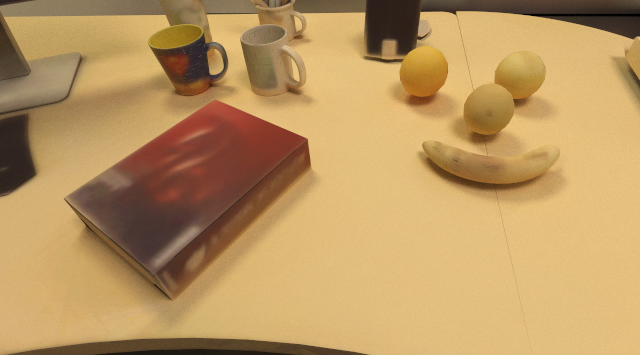} & 
\includegraphics[width=.19\linewidth]{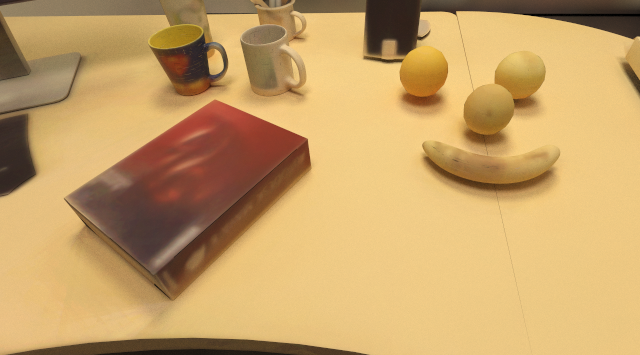} & 
\includegraphics[width=.19\linewidth]{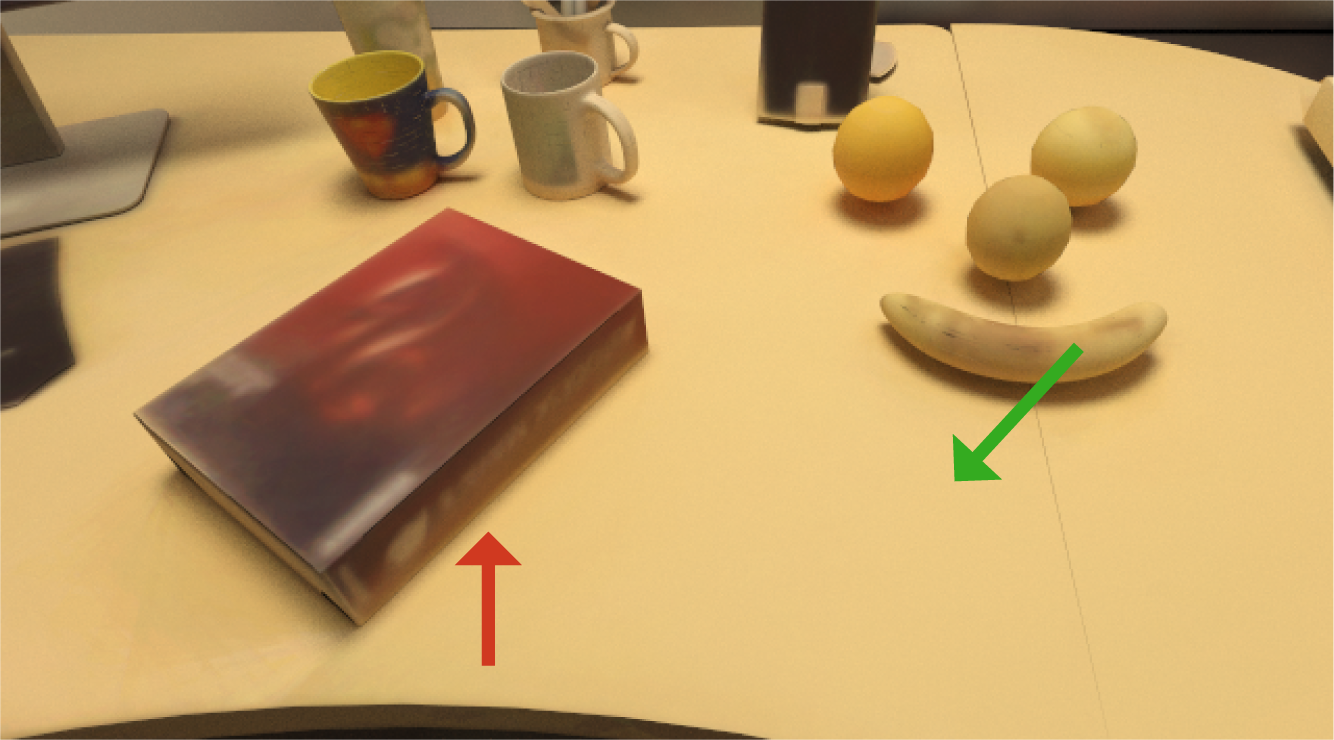} & 
\includegraphics[width=.19\linewidth]{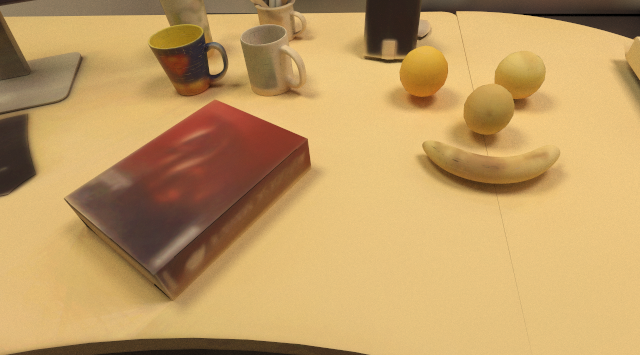} & 
\includegraphics[width=.19\linewidth]{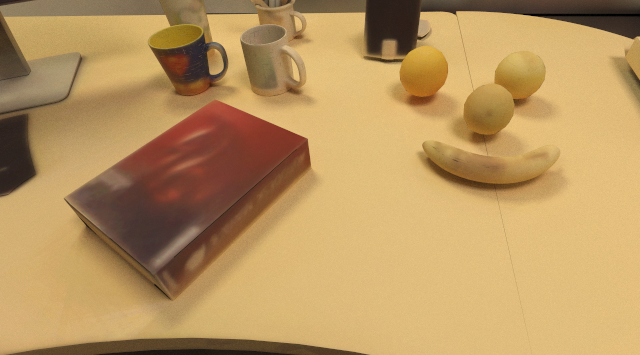} \\ 
{\rotatebox[origin=lB]{90}{\hspace{0.2in}\small{\textsc{Texture}}}} &
\includegraphics[width=.19\linewidth]{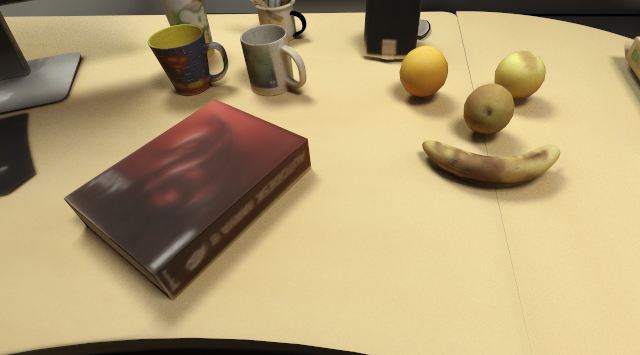} & 
\includegraphics[width=.19\linewidth]{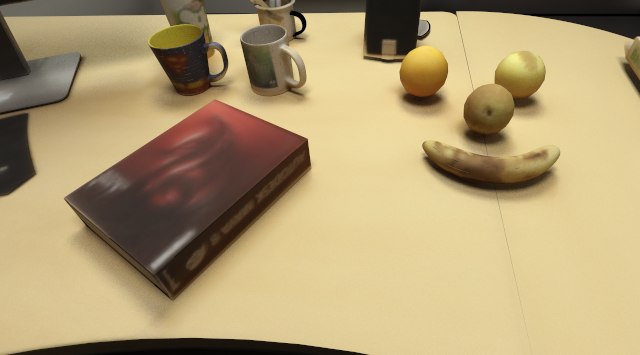} & 
\includegraphics[width=.19\linewidth]{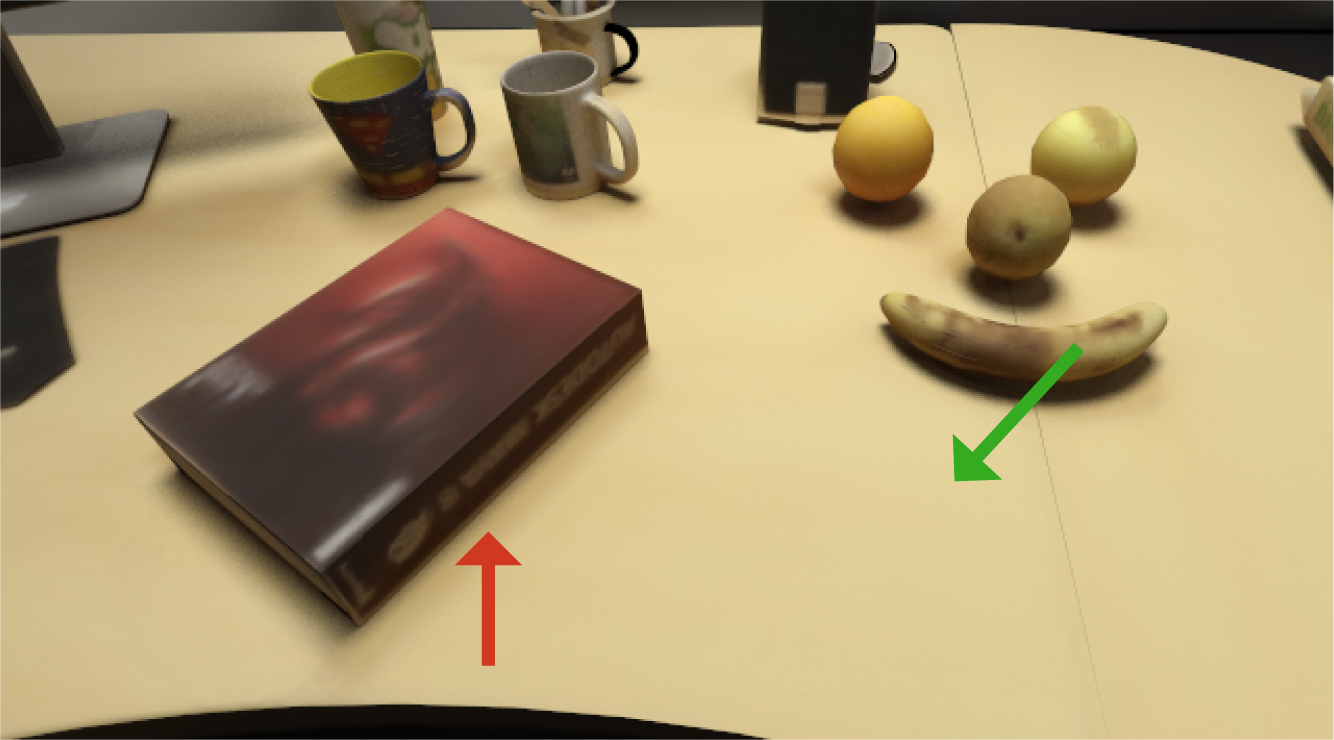} & 
\includegraphics[width=.19\linewidth]{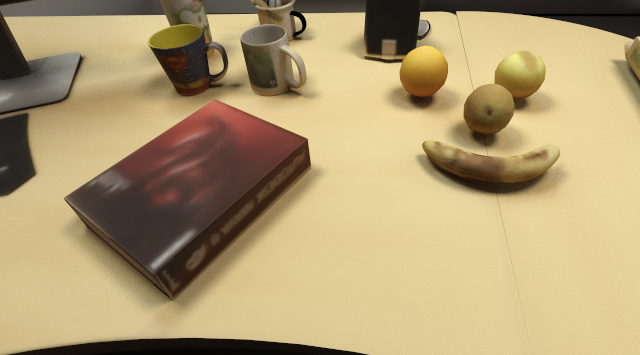} & 
\includegraphics[width=.19\linewidth]{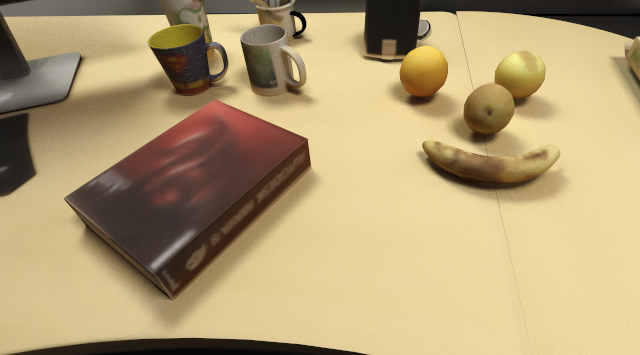} \\ 
%\hline
\end{tabular}
\caption{
\label{fig:comparisons2}
%\TODO{Update \textsc{Li} and annotate.}
Comparison with baselines in \textsc{Real Office} scene. We render a novel viewpoint with changing lighting conditions and show $5$ frames with different lighting conditions. Notice how our method (re)move shadows (red arrow) and reconstruct highlights (green arrow) accurately while both \textsc{LiEtAl} and \textsc{Texture} fail to do so. Please see the supplemental video to appreciate the smooth transition of the highlights and shadows as a result of movement of the light sources.
} 
\end{figure*}

Since we do not have ground truth for real scenes, we show qualitative results on \textsc{Real Office} scene in Figure~\ref{fig:comparisons2}. We show $5$ re-renderings of the scene with modified lighting using our method and the baselines.
In line with our observations on synthetic scenes, we observe that our method is able to predict the variations more accurately than \textsc{LiEtAl} which fails to reconstruct highlights on glossy surfaces such as the tabletop and red box. Compared to \textsc{Texture} we can easily see that the surfaces do not reflect the change in lighting and the baked-in shadows and highlights are still present due to the static and diffuse nature of the texture; please see the video to appreciate the visual importance of this effect.

\begin{table}[!t]
%\setlength{\tabcolsep}{3pt}
%\small{
\centering
\begin{tabular}{l||ccc}

\multicolumn{4}{c}{\textbf{Synthetic Veach Ajar}} \\
\hline \hline
 & \multicolumn{3}{c}{MSE $\downarrow$}    \\
%\hline
		
Method & Diffuse & Roughness & Specular \\
\hline
Ours  & $\textbf{0.022135}$ & $\textbf{0.055563}$ & $\textbf{0.016522}$  \\
Ours Im. & $0.030718$ & $0.064362$ & $0.049969$  \\
%No RP Tex. Space  & $0.197676$ & $0.418624$ & $0.019125$ \\
No RP  & $0.141682$ & $0.372725$ & $0.055459$ \\
		
\hline \hline
\multicolumn{4}{c}{\textbf{Synthetic Dining Room}} \\
\hline \hline
& \multicolumn{3}{c}{MSE $\downarrow$}   \\
%\hline 
Method & Diffuse & Roughness & Specular \\
\hline
Ours  & $\textbf{0.026491}$ & $0.071609$ & $0.028250$  \\
Ours Im. & $0.026642$ & $0.077578$ & $\textbf{0.024123}$  \\
%No RP Tex. Space & $0.05345$ & $0.04616$ & $0.033434$ \\
No RP & $0.053104$ & $\textbf{0.048222}$ & $0.032385$ \\
\hline \hline
\end{tabular}
%}
\caption{
\label{tab:ablations}
Quantitative evaluation for the ablation on the synthetic scenes  on the material maps, with 1) no reprojected statistics, 2) no multi-view merge in texture space. We see an increase in multi-view information helps improve the maps quality and/or consistency across views~(see also Tab.~\protect{\ref{tab:ablations1}} and Fig.~\protect{\ref{fig:ablations1}}).
}
\end{table}

\begin{table}[!t]
\begin{tabular}{l||cc|cc}
%\hline
 & \multicolumn{2}{c|}{\textbf{Veach Ajar}} & \multicolumn{2}{c}{\textbf{Dining Room}}\\
Method & PSNR $\uparrow$ & DSSIM $\downarrow$ & PSNR $\uparrow$ & DSSIM $\downarrow$ \\
\hline \hline
\textsc{ \small{ \textsc{Ours} } } &\small{$\textbf{19.520969}$} & \small{$\textbf{0.190185}$} & \small{$\textbf{19.35363}$} &\small{$\textbf{0.244253}$}\\
\textsc{ \small{ \textsc{No RP} } } &\small{$11.338421$} & \small{$0.413321$} & \small{$17.675568$}& \small{$0.343187$} \\
\hline \hline
\end{tabular}
\caption{
\label{tab:ablations1}
 Quantitative evaluation for the ablation on the synthetic scenes on the re-renderings, with 1) no reprojected statistics, 2) no multi-view merge in texture space. Using multi-view statistics achieves better re-rendering quality than using only a single view~(see also Tab.~\protect{\ref{tab:ablations}} and Fig.~\protect{\ref{fig:ablations1}}).
}
\end{table}

\begin{figure*}[h!]
\setlength{\tabcolsep}{1pt}
\begin{center}
\begin{tabular}{ccccc}
%\hline
%\hline
%& Input & Diffuse & Roughness & Specular\\
%-------------View 1-----------------
%\hline
{\rotatebox[origin=lB]{90}{\hspace{0.3in}\small{Input}}} & 
\includegraphics[width=.2\linewidth]{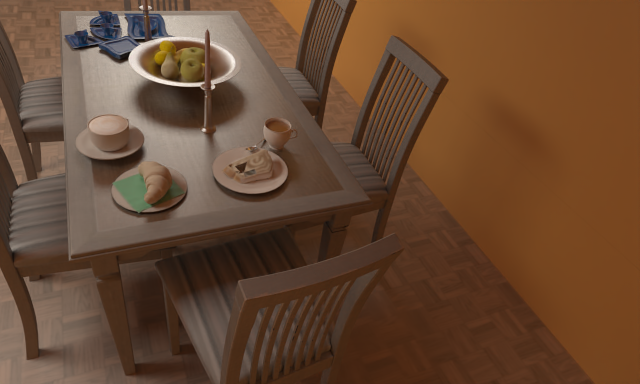} & 
\includegraphics[width=.2\linewidth]{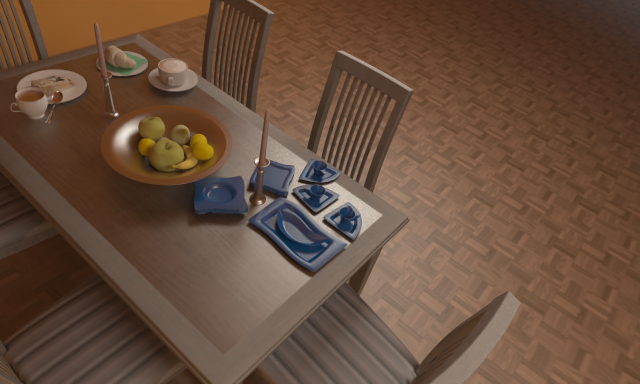} & 
\includegraphics[width=.2\linewidth]{images/ablations/synth_dining/gt/cam96.png} & 
\includegraphics[width=.2\linewidth]{images/ablations/synth_dining/gt/cam129.png} \\

& \multicolumn{2}{c}{Diffuse} & \multicolumn{2}{c}{Roughness} \\

{\rotatebox[origin=lB]{90}{\hspace{0.1in}\small{Ground Truth}}} & 
\includegraphics[width=.2\linewidth]{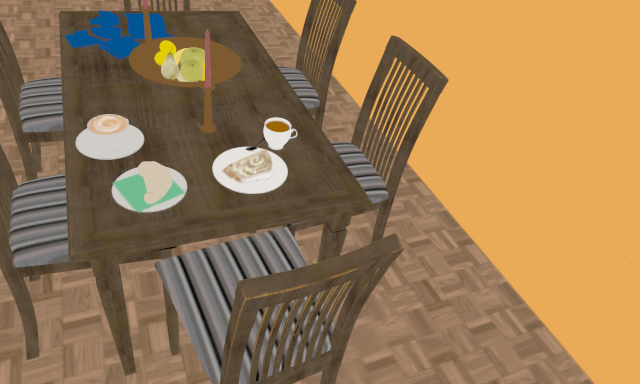} & 
\includegraphics[width=.2\linewidth]{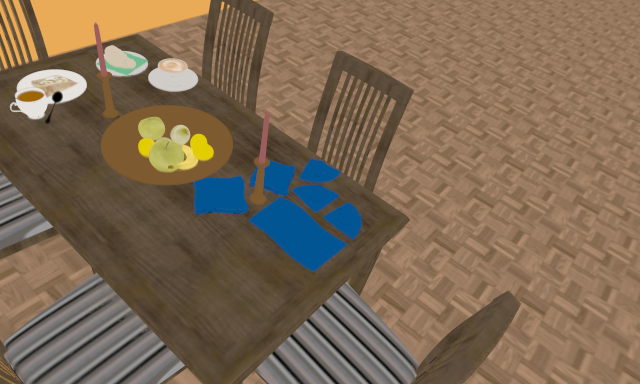} & 
\includegraphics[width=.2\linewidth]{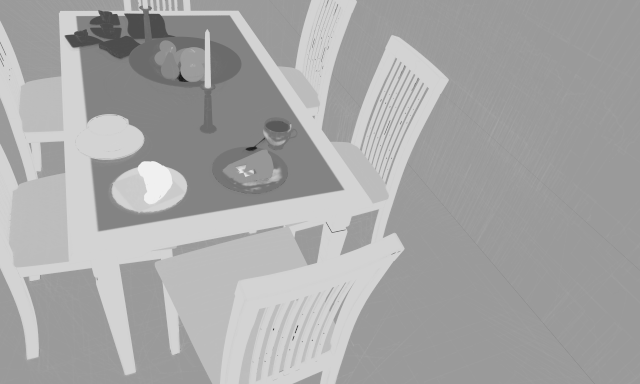} & 
\includegraphics[width=.2\linewidth]{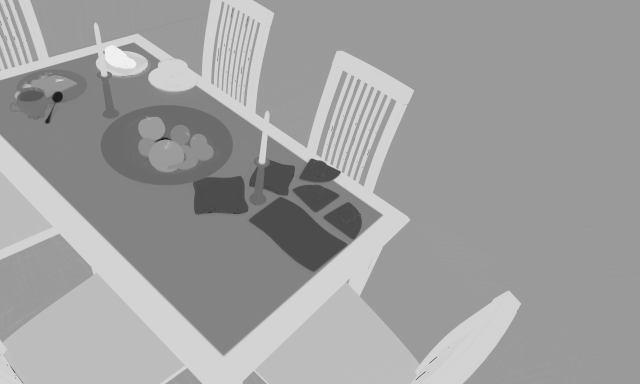} \\

{\rotatebox[origin=lB]{90}{\hspace{0.3in}\small{Ours}}} & 
\includegraphics[width=.2\linewidth]{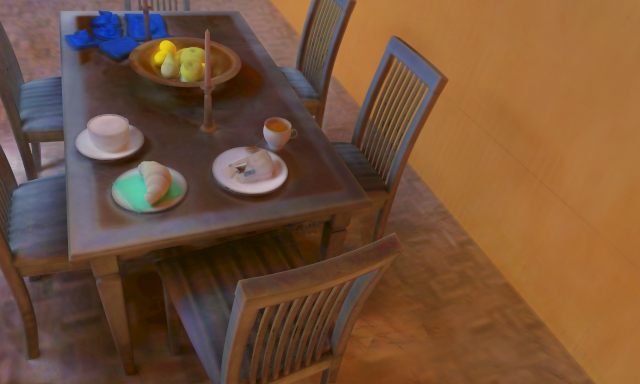}  &
\includegraphics[width=.2\linewidth]{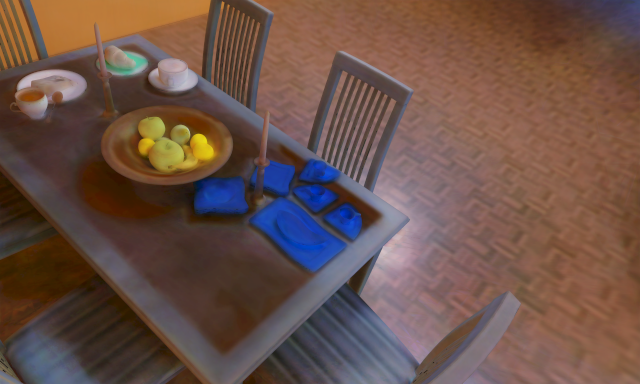}  &
\includegraphics[width=.2\linewidth]{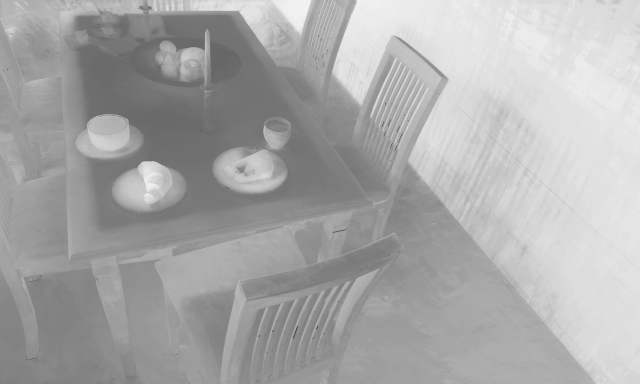} & 
\includegraphics[width=.2\linewidth]{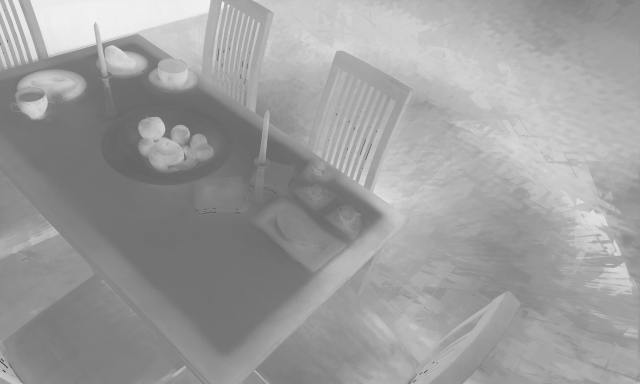} \\

{\rotatebox[origin=lB]{90}{\small{Ours Im. Space}}} & 
\includegraphics[width=.2\linewidth]{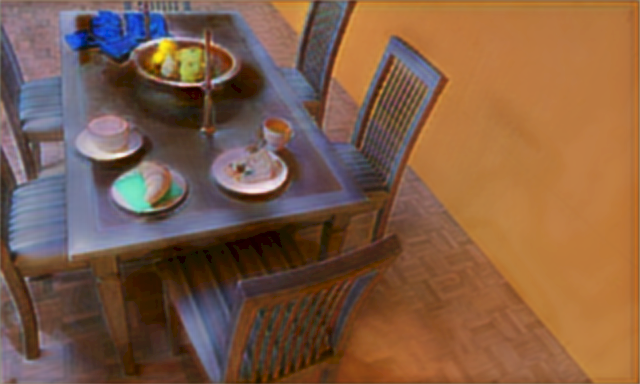} & 
\includegraphics[width=.2\linewidth]{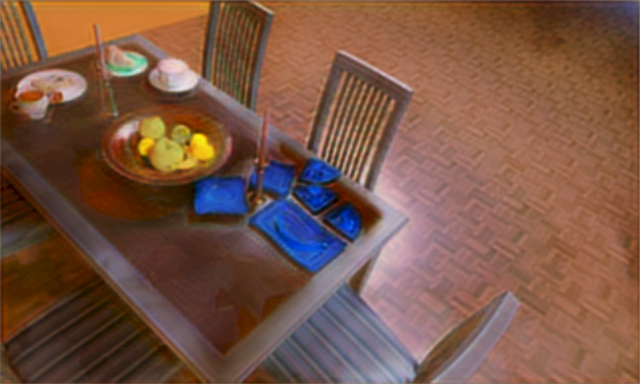} & 
\includegraphics[width=.2\linewidth]{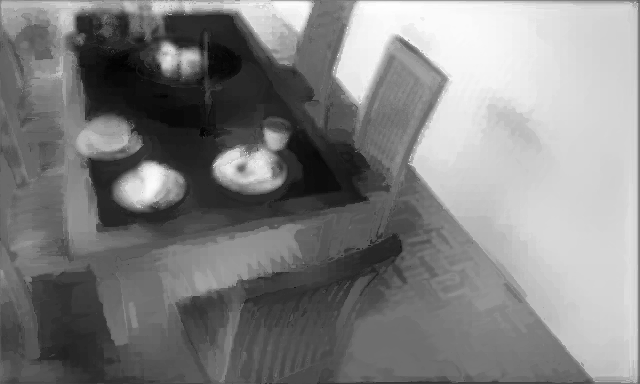} & 
\includegraphics[width=.2\linewidth]{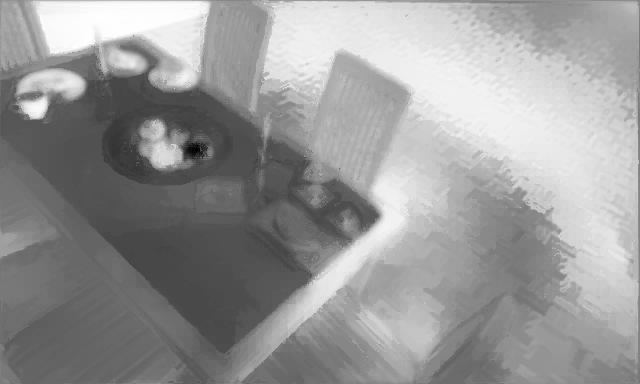} \\ 

{\rotatebox[origin=lB]{90}{\hspace{0.2in}\small{No Reproj.}}} & 
\includegraphics[width=.2\linewidth]{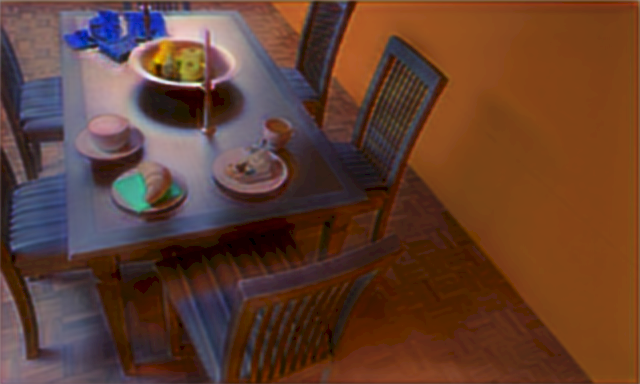} & 
\includegraphics[width=.2\linewidth]{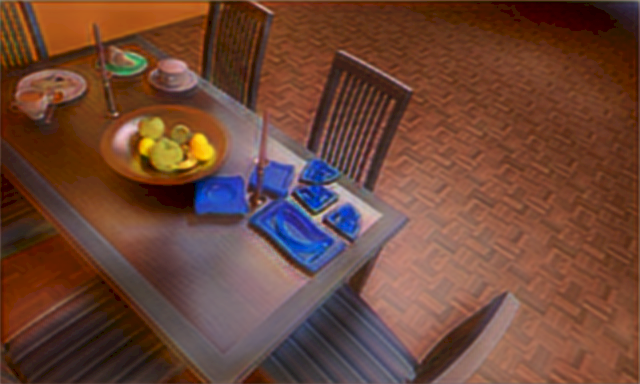} & 
\includegraphics[width=.2\linewidth]{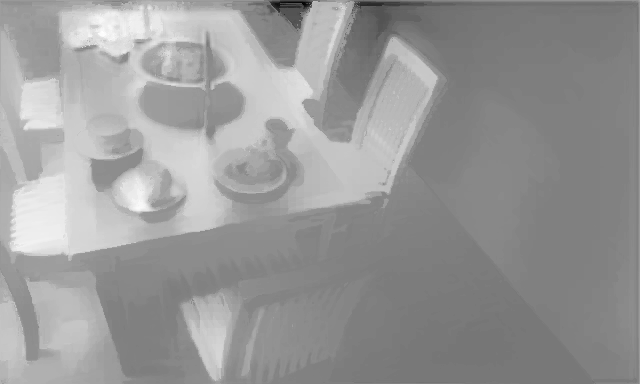} & 
\includegraphics[width=.2\linewidth]{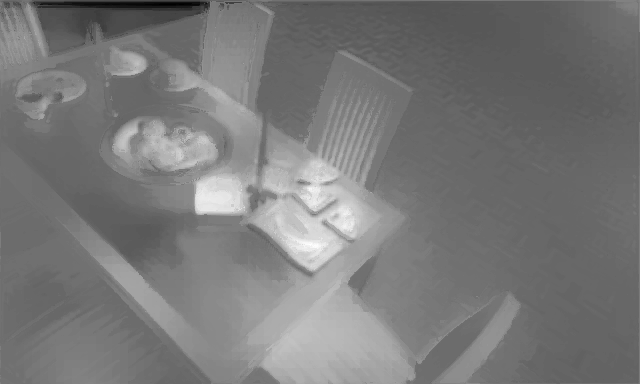} \\

%\hline
& \multicolumn{2}{c}{Specular} & \multicolumn{2}{c}{Re-render} \\
%\hline
{\rotatebox[origin=lB]{90}{\hspace{0.1in}\small{Ground Truth}}} & 
\includegraphics[width=.2\linewidth]{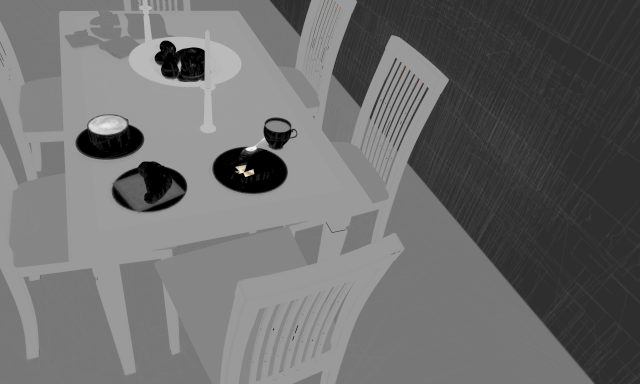}  &
\includegraphics[width=.2\linewidth]{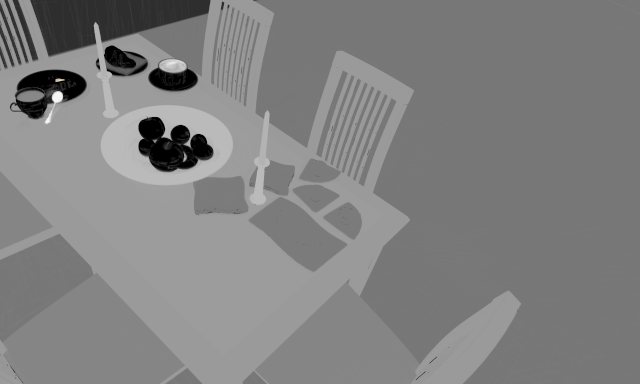}  &
\includegraphics[width=.2\linewidth]{images/ablations/synth_dining/gt/cam96.png}  &
\includegraphics[width=.2\linewidth]{images/ablations/synth_dining/gt/cam129.png}  \\

{\rotatebox[origin=lB]{90}{\hspace{0.3in}\small{Ours}}} & 
\includegraphics[width=.2\linewidth]{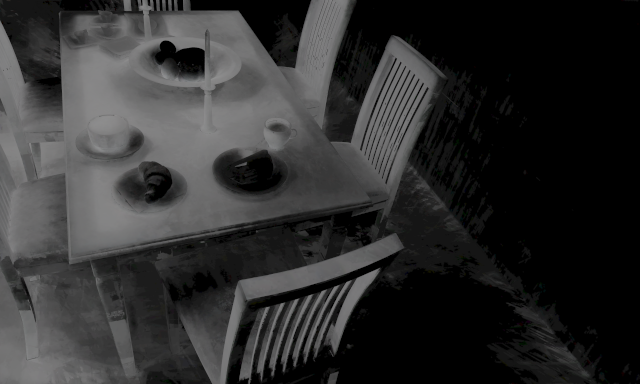} &
\includegraphics[width=.2\linewidth]{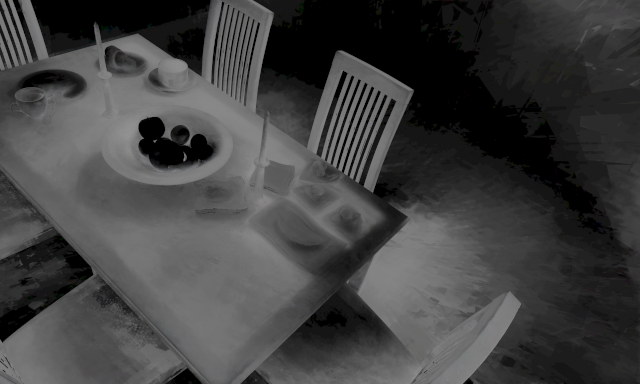} &
\includegraphics[width=.2\linewidth]{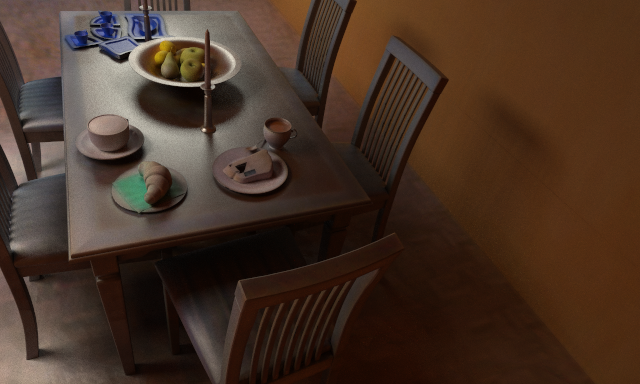} &
\includegraphics[width=.2\linewidth]{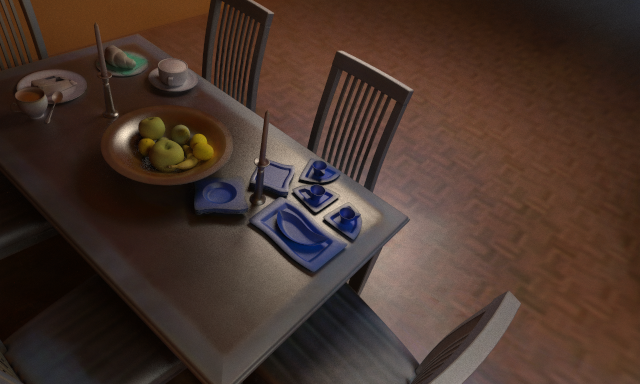} \\

{\rotatebox[origin=lB]{90}{\small{Ours Im. Space}}} &  
\includegraphics[width=.2\linewidth]{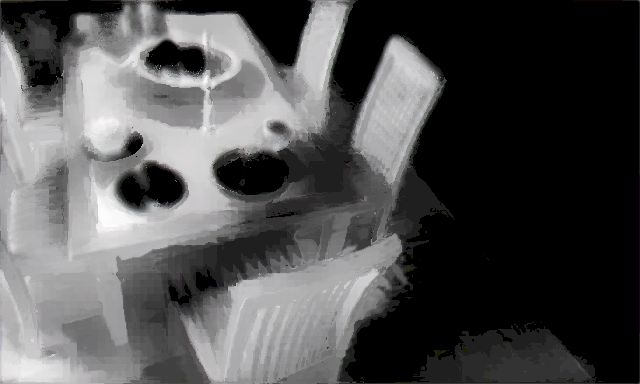} &
\includegraphics[width=.2\linewidth]{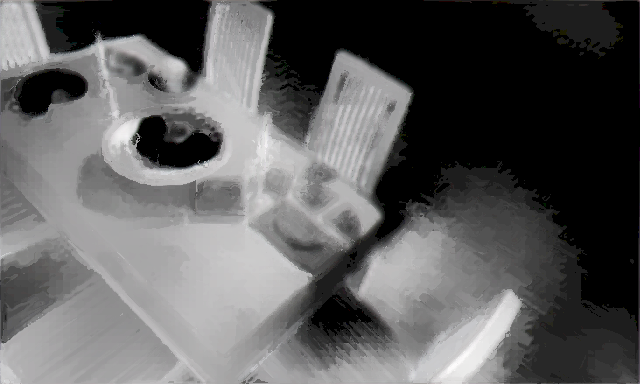} &
 &  \\

{\rotatebox[origin=lB]{90}{\hspace{0.2in}\small{No Reproj.}}} & 
\includegraphics[width=.2\linewidth]{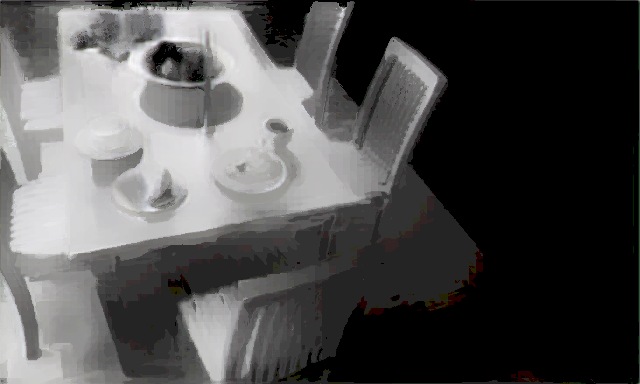} &
\includegraphics[width=.2\linewidth]{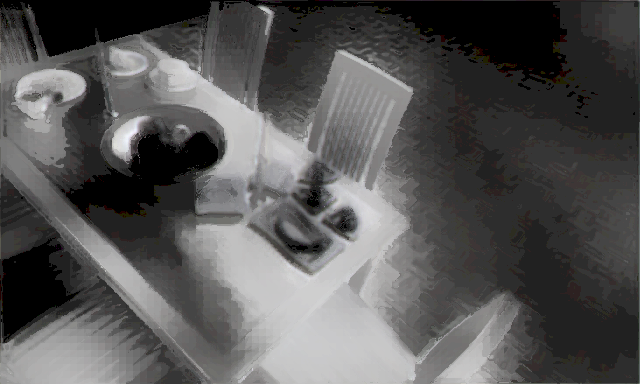} &
\includegraphics[width=.2\linewidth]{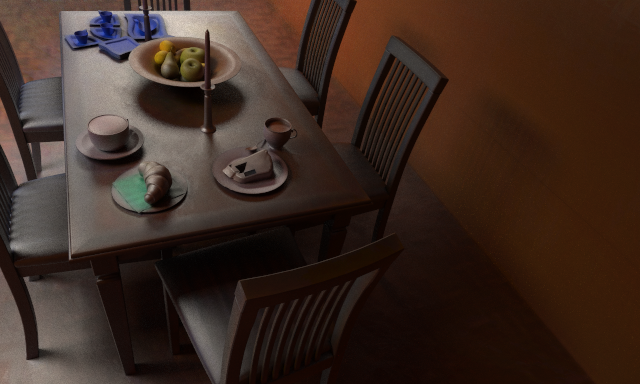}  & 
\includegraphics[width=.2\linewidth]{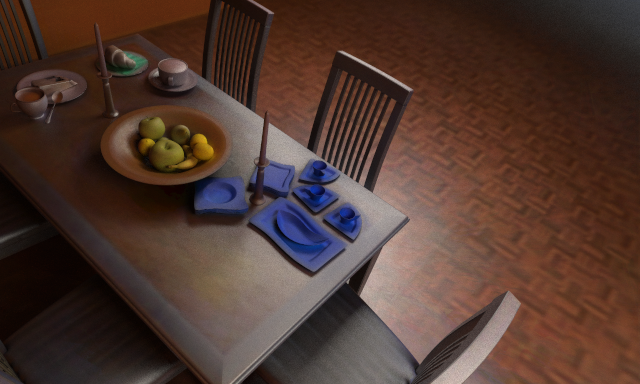}  \\ 
%\hline
\end{tabular}
\end{center}
\caption{
\label{fig:ablations1}
Example images from \textsc{Synthetic Dining Room} showing the effect of increasing multi-view information on results.
Note how the quality of the maps is significantly improved by using the reprojected statistics observed in our image space predicted maps as compared to no reprojection, i.e. using only a single image.
Furthermore, gathering the image space maps in texture space helps improve the consistency of the maps across views and thus improves re-rendering by assigning same material in local regions (esp. in roughness and specular maps). 
As a result, re-rendering is closer to ground truth.
}
\end{figure*}

\subsection{Ablations}

\subsubsection{Multi-view Reprojections}
\label{sec:analysis}
%We performed the following ablations: (a) Joint vs. Individual track training + fine tuning, (b)
%Specular network: With/without additional buffers \GD{SPECIFY}

We perform a first ablation on the two main components of our algorithm: 1) we remove the statistics reprojected from other views (No RP) and 2) we remove the merging of maps in texture space (Ours Im.). We show quantitative comparisons, where we provide error in form of Mean Squared Error (MSE) for the three maps computed for $10$ randomly selected input views.

We show quantitative results in Tab.~\ref{tab:ablations},\ref{tab:ablations1} and an example of the visual effect of the different cases of increasing multi-view information with each step
%\GD{algorithmic} choices \GD{we made} 
in Fig.~\ref{fig:ablations1}. 
Using all our components improves results in the majority of cases. 
%We show visual results in Fig.~\ref{fig:ablations1} for the Dining Room scene. 
The use of the reprojected image statistics makes a very significant difference in the quality of the maps. 
While merging in texture space may not significantly improve quality, it helps with increasing consistency between different views for the underlying surface material properties (especially for roughness and specular maps) and thus helps improve quality of the final re-rendering. Reprojection improves roughness on most of the objects in the scene, but sometimes has a negative effect on background parts that lack observations (see Fig.~\ref{fig:ablations1}); this explains why No-RP has better MSE for roughness in Dining Room in Tab.~\ref{tab:ablations}. 
% \GD{REMOVE? The use of Spherical Gaussian lights is less pronounced, but does help slightly in the estimation of the specular map, which in turn has a small effect on the intensity of highlights.}

\subsubsection{Inaccurate Geometry}

\begin{figure}[!t]
\setlength{\tabcolsep}{1pt}
\begin{tabular}{ccc}
%\hline
 & MVS Mesh & Re-topologized Mesh \\
%\hline
{\rotatebox[origin=lB]{90}{\hspace{0.2in}\small{Re-render}}} & 
\includegraphics[width=.48\linewidth]{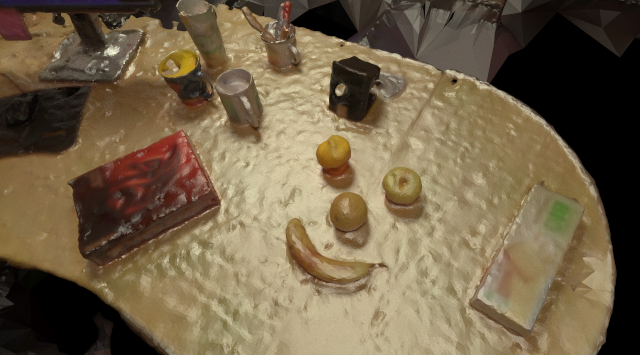} &
\includegraphics[width=.48\linewidth]{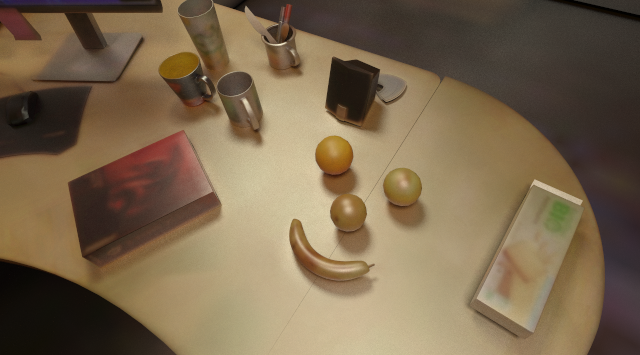} \\ 

{\rotatebox[origin=lB]{90}{\hspace{0.2in}\small{Albedo}}} & 
\includegraphics[width=.48\linewidth]{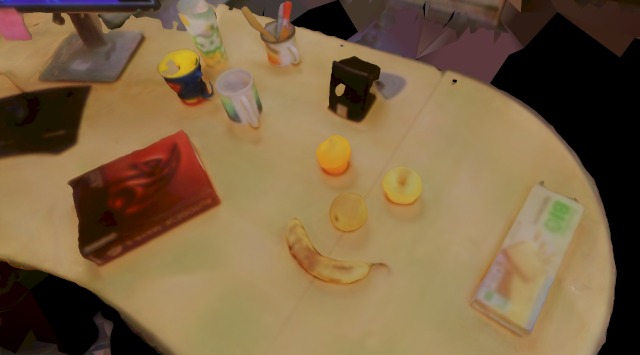} &
\includegraphics[width=.48\linewidth]{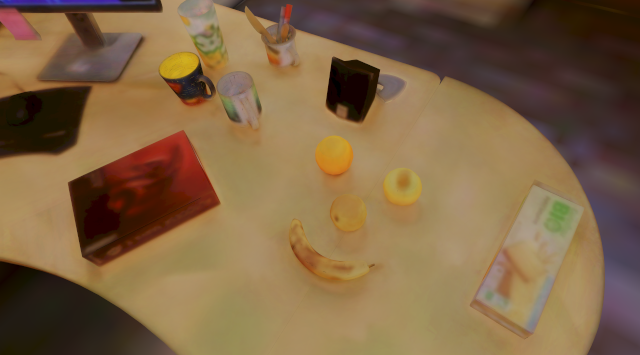} \\ 

{\rotatebox[origin=lB]{90}{\hspace{0.2in}\small{Roughness}}} & 
\includegraphics[width=.48\linewidth]{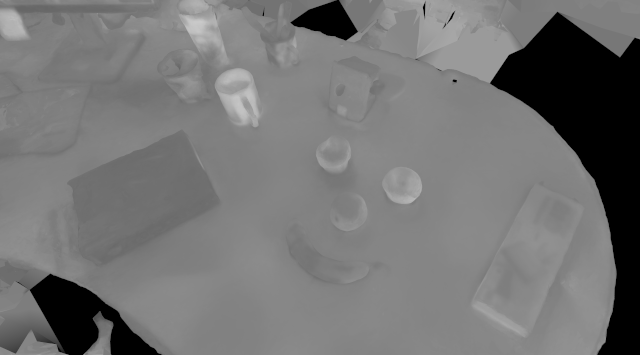} &
\includegraphics[width=.48\linewidth]{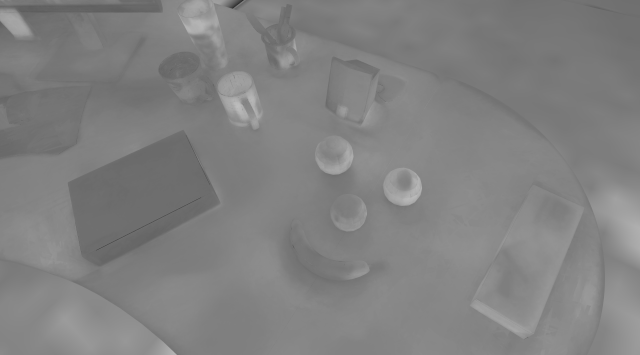} \\ 

{\rotatebox[origin=lB]{90}{\hspace{0.2in}\small{Specular}}} & 
\includegraphics[width=.48\linewidth]{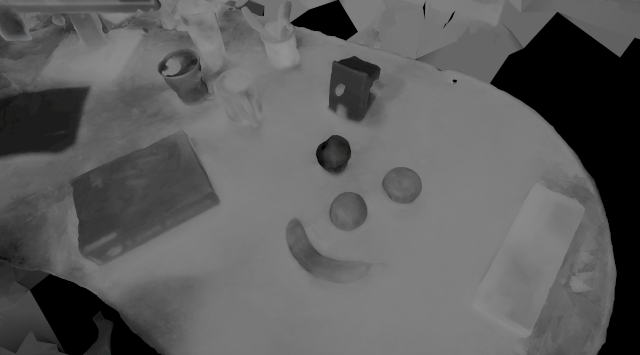} &
\includegraphics[width=.48\linewidth]{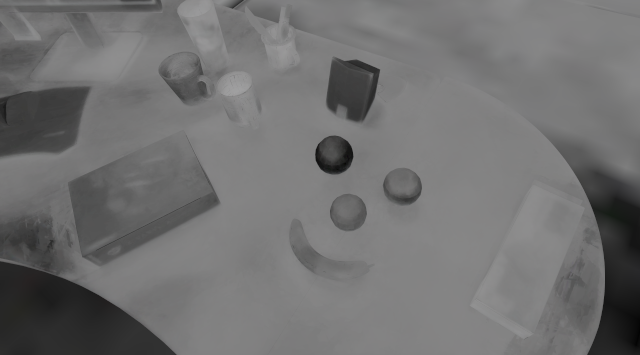} \\ 

%\hline
\end{tabular}
\caption{
\label{fig:mvs}
%\TODO{Update \textsc{Li} and annotate.}
Effect of inaccurate geometry. 
We run our pipeline on the MVS mesh obtained directly from~\cite{reality2016capture}. The results from the MVS mesh is on the left column and our re-topologized mesh on the right column.
We show a re-rendering in the first row, followed by the albedo, roughness and specular maps obtained using our method in subsequent rows respectively.}
 
\end{figure}

We study the robustness of our method for inaccurate geometry by running our pipeline on the degraded mesh obtained directly from multi-view stereo (MVS)~\cite{reality2016capture} which consists of large holes and bumpy surfaces. 
\NEW{The bumpy surfaces are an instance of extreme vertex perturbation.}
We show a qualitative comparison in Fig.~\ref{fig:mvs}.
From the figure we can confirm that the maps obtained from the MVS mesh is only slightly degraded compared to the re-topologized mesh. 
Thus, our material estimation is robust to geometrical inaccuracies.
While the maps obtained are similar, the final image obtained after re-rendering is highly degraded when rendered since the MVS geometry has bumps and holes.
To obtain high quality re-renderings we need good geometry, justifying our design choice of using re-topology. In future work, it may be possible to adapt previous methods (e.g., ~\cite{yu_cvpr22, Bauchet_tog20}) to provide geometry that corrects these errors for flat surfaces, but it would be necessary to preserve the relatively well reconstructed irregular objects (such as the fruit on the table).

%% file: 6_conclusion.tex
\section{Limitations and Future Work}

\begin{wrapfigure}[7]{r}{0.35\linewidth}
\vspace{-0.17in}
\includegraphics[width=\linewidth]{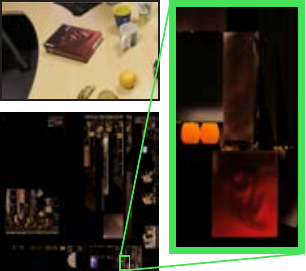}
\end{wrapfigure}
Despite yielding convincing re-renderings on synthetic and on real scenes, our method still has several limitations.
Our final texture-space maps often suffer from limited resolution, since the texture atlas can provide only limited space for a given object (see inset, where the texture of the red box covers only a very small part of the texture atlas). 
This is an inherent problem with texture-space methods, and alternative approaches (e.g., see~\cite{yuksel2019rethinking}) could be a good direction for future work.
While reprojection error may be a contributing factor for blurriness in results, we believe our use of median filter to merge and obtain texture-space maps alleviates this problem.
%\paragraph*{Normal mapping.}
We focused on estimating BRDF parameters for each texel of a texture atlas. A natural extension would be to also estimate per-texel normals expressed in a local coordinate frame, which would allow the reproduction of small geometric details not modeled in the retopologized mesh. However, such small-scale relief is often difficult to perceive when captured from far away.
%\paragraph*{Training data.}

\setlength{\columnsep}{0.08in}
\begin{wrapfigure}[7]{r}{0.3\linewidth}
\vspace{-0.1in}
\includegraphics[width=\linewidth]{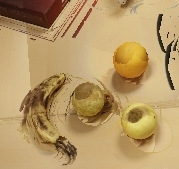}
\end{wrapfigure}

In some cases, e.g., the banana and pear, our method predicts glossiness that is high; we hypothesize that this is due to the re-projection errors due to the mismatch between the re-topologized and real scene geometry. The effect of this is visible in the reprojected min image (see inset).
%\SP{I feel this should be in limitations, but I get the point of being up front about the results. My only worry is it distracts the reviewers from positive results here and paint our results in a negative light more than is needed at this moment. I'll keep it here for now. }

The dataset we created to demonstrate our approach offers limited variability, which in turns limits the ability of our method to handle diverse scenes. While we provide our toolbox to generate additional training images, rendering large datasets is costly and could benefit from strategies to reuse computation across views \cite{Fraboni2019}; we hypothesize that augmenting the variety and the number of training images seen by our network will improve results overall, possibly helping remove shadow and incorrect color residuals that are sometimes still present in our albedo maps.

%\paragraph*{Fixed vs. learned operations.}
We rely on fixed color statistics to aggregate the multi-view information that we feed to our per-view CNNs, and we employ a fixed median filter to merge the resulting per-view predictions into texture space. Replacing these two operations by differentiable pooling in a learned feature space could yield improved predictions, as has been done in other applications~\cite{su15mvcnn,Kalogerakis2017}. 
%in other multi-view architectures for object classification \cite{su15mvcnn} and segmentation \cite{Kalogerakis2017}. 
However, training such an architecture end-to-end raises specific challenges, such as storing multiple CNN tracks in memory and performing differentiable re-projection in the texture atlas while doing per-image processing.

\section{Conclusion}
We have presented the first attempt at creating scene-scale material map textures of indoor environments using deep learning.
Our solution retains the strength of image-space CNNs, which have proven successful at recovering material parameters for close-up photographs of flat surfaces and isolated objects.
To apply such CNNs at scene scale, we first inject multi-view information by computing statistics about a pixel color when re-projected into neighboring views. 
We then move to texture space to merge per-view predictions into a single texture atlas suitable for rendering.

Our method allows automatic generation of material maps that allow plausible renderings, retaining the overall look of multiple objects in a scene, both for synthetic scenes with available ground truth and for real scenes. 
Our results demonstrate that by exploiting reprojected multi-view data, improving the rendering loss and exploiting state-of-the-art components it is possible to provide an operational pipeline to extract convincing materials at scene-scale from a set of images as input.

%% file: acknowledgement.tex
\section*{Acknowledgments}
The research was funded by the ERC Advanced grant FUNGRAPH No 788065 (\url{https://project.inria.fr/fungraph/}).
The authors are grateful to Inria Sophia Antipolis - M\'editerran\'ee ``Nef" computation cluster and the OPAL infrastructure from Universit\'e C\^ote d'Azur for providing resources and support.
We thank the associate editor and the anonymous reviewers for their insightful comments which helped improve the manuscript.
We also thank B. Bitterli for the scenes in his rendering repository.
The authors also extend their thanks to Felix H{\"a}hnlein and Emilie Yu for their helpful disucssion and support for the project and especially thank the 3D artist Stefania Kousoula for mesh refinement and re-topology. 